\documentclass[amsmath,amssymb,apj]{emulateapj-rtx4}
\usepackage{color}

\begin{document}


\title{Constraint on the ejecta mass for a black hole-neutron star merger event candidate S190814\lowercase{bv}}


\author{Kyohei Kawaguchi}
\affil{Institute for Cosmic Ray Research, The University of Tokyo, 5-1-5 Kashiwanoha, Kashiwa, Chiba 277-8582, Japan}
\affiliation{Center for Gravitational Physics,
  Yukawa Institute for Theoretical Physics, 
Kyoto University, Kyoto, 606-8502, Japan} 
\author{Masaru Shibata}
\affil{Max Planck Institute for Gravitational Physics (Albert Einstein Institute), Am M\"{u}hlenberg 1, Potsdam-Golm, 14476, Germany}
\affiliation{Center for Gravitational Physics,
  Yukawa Institute for Theoretical Physics, 
Kyoto University, Kyoto, 606-8502, Japan} 

\and
\author{Masaomi Tanaka}
\affil{Astronomical Institute, Tohoku University, Aoba, Sendai 980-8578, Japan}

\newcommand{\angstrom}{\text{\normalfont\AA}}
\newcommand{\rednote}[1]{{\color{red} (#1)}}
\newcommand{\UTF}[1]{\rednote{UTF[#1]}}

\begin{abstract}
We derive the upper limit to the ejecta mass of S190814bv, a black hole-neutron star merger candidate,  through the radiative transfer simulations for kilonovae with the realistic ejecta density profile as well as the detailed opacity and heating rate models. The limits to the ejecta mass strongly depend on the viewing angle. For the face-on observations ($\le45^\circ$), the total ejecta mass should be smaller than $0.1\,M_\odot$ for the average distance of S190814bv ($D=267$ Mpc), while larger mass is allowed for the edge-on observations. We also derive the conservative upper limits of the dynamical ejecta mass to be $0.02\,M_\odot$, $0.03\,M_\odot$, and $0.05\,M_\odot$ for the viewing angle $\le 20^\circ$, $\le 50^\circ$, and for $\le 90^\circ$, respectively. We show that the {\it iz}-band observation deeper than $22$ mag within 2 d after the GW trigger is crucial to detect the kilonova with the total ejecta mass of $0.06\,M_\odot$ at the distance of $D=300$ Mpc. We also show that a strong constraint on the NS mass-radius relation can be obtained if the future observations put the upper limit of $0.03\,M_\odot$ to the dynamical ejecta mass for a BH-NS event with the chirp mass smaller than $\lesssim 3\,M_\odot$ and effective spin larger than $\gtrsim 0.5$.
\end{abstract}


\section{Introduction}
A black hole-neutron star (BH-NS) merger, which is one of the main targets of ground-based gravitational-wave (GW) detectors~(LIGO:~\citealt{TheLIGOScientific:2014jea}, Virgo:~\citealt{TheVirgo:2014hva}, KAGRA:~\citealt{Kuroda:2010zzb}), can be accompanied with an electromagnetic (EM) counterpart if the NS is tidally disrupted~\citep{1991AcA....41..257P,Li:1998bw}. At the onset of tidal disruption, a part of NS material would be ejected from the system (referred to as the dynamical ejecta)~\citep{Rosswog:2005su,Shibata:2007zm,Etienne:2008re,Lovelace:2013vma,Foucart:2014nda,Kyutoku:2015gda,Kawaguchi:2015bwa,Kawaguchi:2016ana,Kyutoku:2017voj,Foucart:2019bxj}. Subsequently, additional ejecta would be launched from the accreting torus around the remnant BH, driven by amplified magnetic fields or effective viscous heating due to magnetic turbulence~\citep{Fernandez:2013tya,Metzger:2014ila,Just:2014fka,Kiuchi:2015qua,Siegel:2017nub,Lippuner:2017bfm,Siegel:2017jug,Ruiz:2018wah,Fernandez:2018kax,Christie:2019lim,Fujibayashi:2020qda} (referred to as the post-merger ejecta). Such ejected material would be the source of the so-called kilonova, which is an EM transient phenomenon of which the emission is powered by radioactive decays of heavy r-process elements synthesized in the ejecta~\citep{Li:1998bw,Kulkarni:2005jw,Metzger:2010sy,Kasen:2013xka,Tanaka:2013ana,Tanaka:2013ixa}. 

Mass of the dynamical and post-merger ejecta can be either higher or lower than those formed in NS-NS merger, depending strongly on the binary parameters, such as the NS mass, BH mass, NS radius, and BH spin~\citep{Rosswog:2005su,Shibata:2007zm,Etienne:2008re,Lovelace:2013vma,Kyutoku:2015gda,Foucart:2018rjc}. It is worth noting that tidal disruption of the NS does not always occur for a BH-NS merger particularly for the cases that the BH mass is large or BH spin is small or NS radius is small~(see e.g.,~\cite{Foucart:2018rjc}). For such a case, no EM counterparts (emitted after merger) will accompany with the detection of GWs from a BH-NS merger (however see e.g.,~\cite{Carrasco:2020sxg,Most:2020ami} for the possible EM precursors). Thus, the detection or the non-detection of the kilonova from a BH-NS merger provides us with important information of the binary parameters complementary to that inferred by the GW data analysis.

On 2019 August 14 advanced LIGO and advanced Virgo have reported the detection of GWs from a BH-NS merger, which is referred to as S190814bv, with a significantly low value of false alarm rate ($10^{-25}\,{\rm yr.}^{-1}$)~\citep{gcn25333}. The luminosity distance to the event is estimated to be $D=267\pm52$ Mpc ($1\sigma$) and the sky localization is achieved within the area of 23 (5) ${\rm deg}^2$ for 90\% (50\%) confidence. EM follow-up observations were performed by many groups~\citep[e.g.,][]{gcn25381,gcn25417,Gomez:2019tuj,Andreoni:2019qgh,Ackley:2020qkz}. Although no significant EM counterpart was found, upper limits in the nearly whole sky localization region of the event are obtained by their efforts.

The upper limit to the ejecta mass is discussed in~\cite{Gomez:2019tuj},~\cite{Andreoni:2019qgh}, and~\cite{Ackley:2020qkz} based on the upper limits to the EM counterparts. By employing an 1d analytical model of~\cite{Villar:2017wcc},~\cite{Gomez:2019tuj} explore the ranges of ejecta mass, velocity, and opacity in which the kilonova emission is consistent with the upper limits obtained by their observation (see Figure 4 in~\cite{Gomez:2019tuj}).~\cite{Andreoni:2019qgh} employ an 1d kilonova model of~\cite{Hotokezaka:2019uwo} and a 2d kilonova model of~\cite{Bulla:2019muo,Dhawan:2019phb} and suggest that the total ejecta mass should be less than $0.04\,M_\odot$ for the face-on observation or less than $0.03\,M_\odot$ for the ejecta opacity $\lesssim2\,{\rm cm^2/g}$ for $D=215\,{\rm Mpc}$.~\cite{Ackley:2020qkz} employ a 2d kilonova model of~\cite{Barbieri:2019sjc} and show that the total ejecta mass larger than $0.1\,M_\odot$ is excluded with high confidence. However, there are several remarks for the kilonova models employed in the previous work. For the 1d model of~\cite{Hotokezaka:2019uwo} and the 2d model of~\cite{Barbieri:2019sjc}, simplified semi-analytical models are employed for radiative transfer with a constant gray opacity. For the 2d radiative transfer model~\cite{Bulla:2019muo,Dhawan:2019phb}, the ejecta model with simplified geometry and heating rate are employed, and the temperature and opacity evolution is given a priori. The temperature and opacity evolution of the ejecta as well as the radiative transfer effect between the multiple ejecta components with non-spherical geometry are crucial for the quantitative prediction of the kilonova lightcurves~\citep{Kawaguchi:2018ptg,Bulla:2019muo,Kawaguchi:2019nju,Darbha:2020lhz}. Thus, while these previous models may give a semi-quantitative idea for the constraint, an independent quantitative analysis deserves to be performed in a wide range of ejecta parameter space.

In this paper we report our study for constraining the ejecta mass of S190814bv by performing the radiative transfer simulations for BH-NS kilonovae with the detailed opacity and heating rate models. In this study, kilonova lightcurves are calculated by employing the ejecta model motivated by numerical-relativity simulations~\citep[e.g.,][]{Foucart:2014nda,Kyutoku:2015gda,Foucart:2015vpa,Kyutoku:2017voj,Foucart:2019bxj,Metzger:2014ila,Wu:2016pnw,Siegel:2017nub,Siegel:2017jug,Fernandez:2018kax,Christie:2019lim,Fujibayashi:2020qda} and by systematically varying the mass of ejecta components. This paper is organized as follows: The setups for the radiative transfer simulation and the ejecta model employed in this work are described in Section~\ref{sec:method}. In Section~\ref{sec:results}, we show the upper limits to the ejecta mass of S190814bv. We compare our results with the previous studies by~\cite{Andreoni:2019qgh} and~\cite{Ackley:2020qkz} in Section~\ref{sec:discussion}. Implications to the future observation are presented in Section~\ref{sec:discussion}. An idea to constrain the NS mass-radius relation by joint analysis employing the upper limit to the ejecta mass with the GW parameter estimation is also discussed in Section~\ref{sec:discussion}. We summarize this work in Section~\ref{sec:summary}. Throughout the paper, magnitudes are given in the AB magnitude system.

\section{Method}\label{sec:method}

\subsection{Radiative Transfer simulation}

We calculate the light curves of kilonova models for BH-NS mergers by a wavelength-dependent radiative transfer simulation code~\citep{Tanaka:2013ana,Tanaka:2017qxj,Tanaka:2017lxb,Kawaguchi:2019nju}. The photon transfer is calculated by a Monte Carlo method for given ejecta profiles of density, velocity, and element abundance. The nuclear heating rates are determined by employing the results of r-process nucleosynthesis calculations by~\cite{Wanajo:2014wha}. We also consider the time-dependent thermalization efficiency following an analytic formula derived by~\cite{Barnes:2016umi}. Axisymmetry is imposed for the matter profile, such as the density, temperature, and abundance distribution. The ionization and excitation states are calculated under the assumption of local thermodynamic equilibrium (LTE) by using the Saha ionization and Boltzmann excitation equations. Special-relativistic effects on photon transfer and light travel time effects are fully taken into account.

For photon-matter interaction, we consider bound-bound, bound-free, and free-free transitions and electron scattering for a transfer of optical and infrared photons~\citep{Tanaka:2013ana,Tanaka:2017qxj,Tanaka:2017lxb}. The formalism of the expansion opacity~\citep{1993ApJ...412..731E,Kasen:2006ce} and the updated line list calculated in~\cite{Tanaka:2019iqp} are employed for the bound-bound transitions. The line list is constructed by an atomic structure calculation for the elements from $Z=26$ to $Z=92$, and supplemented by Kurucz's line list for $Z < 26$~\citep{1995all..book.....K}, where $Z$ is the atomic number. In particular, we restrict the line list for the transitions of which ${\rm ln}(g_l f_l)$ is larger than $-2.5$ to reduce the computational cost, where $g_l$ and $f_l$ denote the statistical weight and the oscillator strength of the transition, respectively. By this prescription, the line list includes $\approx 7\times10^{6}$ lines. We find that the {\it griz}-band emission obtained by employing this restricted line list results to be uniformly brighter by $\approx 0.2\,{\rm mag}$ than those employing the line list with ${\rm ln}(g_l f_l)>-3$. While we should note that uncertainties in the opacity table, heating rate, and the ejecta profile could be larger than this prescription, the brightness of the model lightcurves shown in this paper is reduced by $0.2\,{\rm mag}$ to correct this effect.

\subsection{Ejecta profile}
\begin{figure}
 	 \includegraphics[width=1.1\linewidth]{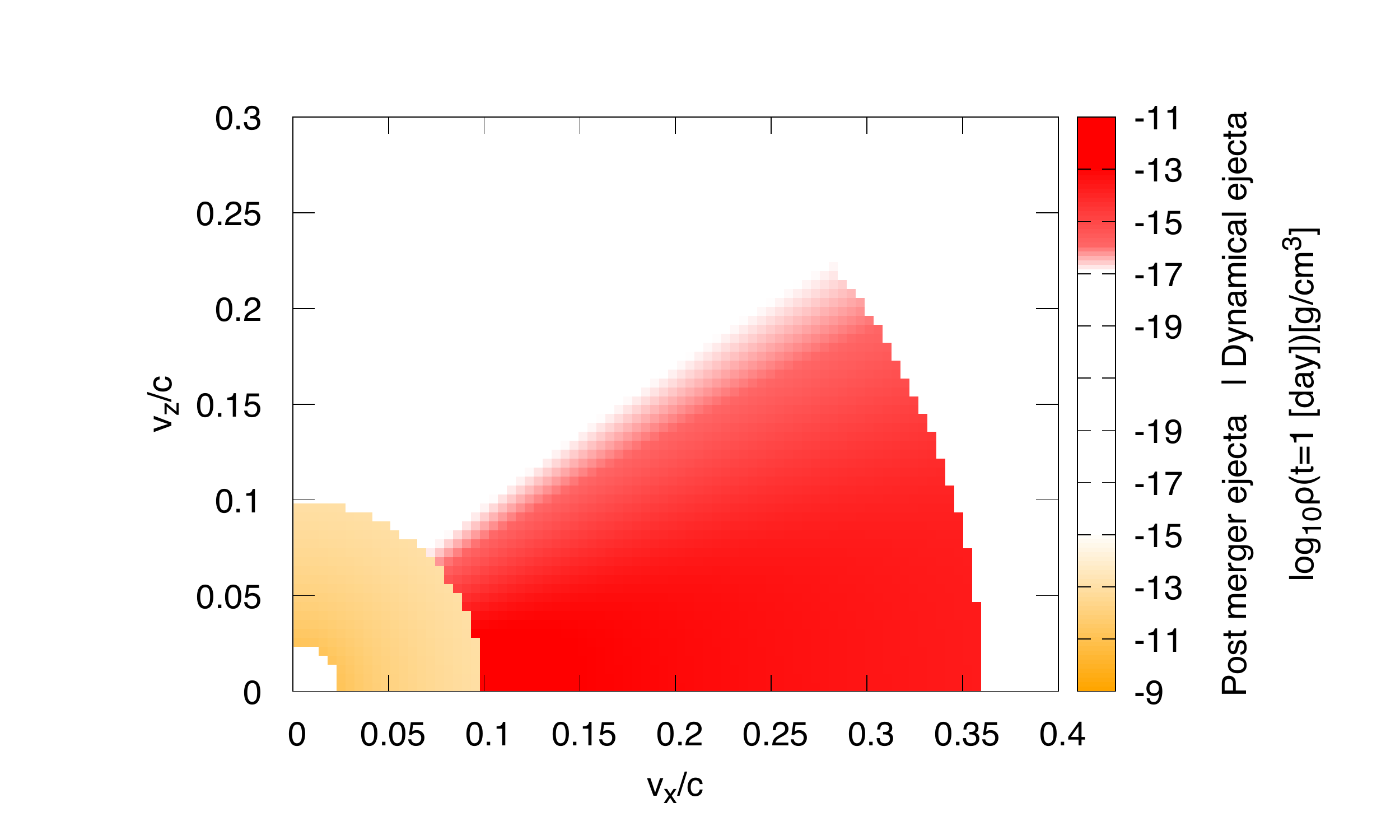}
 	 \caption{Ejecta density profile employed in the radiative transfer simulation. The density profile for  $M_{\rm d}=M_{\rm pm}=0.02\,M_\odot$ is shown as an example. The red and orange regions denote the dynamical and post-merger ejecta, respectively. Homologous expansion of the ejecta and axisymmetry around the rotational axis ($z$-axis) are assumed in the simulation.}
	 \label{fig:dens}
\end{figure}
Axisymmetric homologously expanding ejecta models that consist of the dynamical ejecta with non-spherical geometry and the post-merger ejecta with spherical geometry are employed in this work (see Figure~\ref{fig:dens}).\footnote{We note that the dynamical ejecta could exhibit non-axisymmetric morphology in reality~\citep{Foucart:2014nda,Kyutoku:2015gda}.} As in our previous study~\citep{Kawaguchi:2019nju}, we employ the following density profile for the BH-NS ejecta models motivated by the results of numerical-relativity simulations~\citep{Foucart:2014nda,Kyutoku:2015gda,Siegel:2017nub,Siegel:2017jug,Fernandez:2018kax,Christie:2019lim,Fujibayashi:2020qda}:
\begin{equation}
	\rho\propto
	\left\{\begin{array}{cc}
		r^{-3} &, 0.025\,c\le r/t \le 0.1\,c\\
		{\tilde \Theta}\left(\theta\right)r^{-2} &, 0.1\,c\le r/t\le 0.36\,c\label{eq:dens}
	\end{array}
	\right.,\label{eq:dens_bhns}
\end{equation}
where ${\tilde \Theta}\left(\theta\right)$ is given by
\begin{equation}
	{\tilde \Theta}(\theta)=\frac{1}{1+{\rm exp}\left[-20\left(\theta-1.2\,{\rm [rad]}\right)\right]},
\end{equation}
and $\theta$ is the angle measured from the axis of symmetry. In this model, the dynamical and post-merger ejecta distribute from $0.1\,c$ to $0.36\,c$ and from $0.025\,c$ to $0.1\,c$, respectively. The normalization of the density profile is determined so that the dynamical ejecta mass and post-merger ejecta mass are set to be the assumed values, $M_{\rm d}$ and $M_{\rm pm}$, respectively. The outer edge of the dynamical ejecta ($r/t=v_{\rm d,max}=0.36\,c$) is determined from the condition that its average velocity defined by $v_{\rm d, ave}=\sqrt{2E_{\rm K,d}/M_{\rm d}}$ is $0.25\,c$ with $E_{\rm K,d}$ the kinetic energy of the dynamical ejecta~\citep{Foucart:2014nda,Kyutoku:2015gda}. The average velocity of the post-merger ejecta is set to be $0.06\,c$ following the results of numerical-relativity simulations~\citep[e.g.,][]{Metzger:2014ila,Siegel:2017nub,Siegel:2017jug,Christie:2019lim,Fujibayashi:2020qda}\footnote{We note that the significant amount of ejecta of which velocity is higher than $0.1\,c$ could be formed in the presence of globally coherent and strong poloidal magnetic fields, although it is not very clear how such magnetic fields are established soon after the onset of merger~\citep{Siegel:2017jug,Christie:2019lim}.}. 

For BH-NS mergers,  collisional shock heating or neutrino irradiation in the merger remnant is weak in contrast to NS-NS mergers~\citep[e.g.,][]{Fujibayashi:2017puw}, and hence, substantial amount of the ejecta components could have low $Y_e$ values. Taking the prediction obtained by numerical simulations into account~\citep{Rosswog:2012wb,Just:2014fka,Foucart:2014nda,Kyutoku:2015gda,Foucart:2015vpa,Wu:2016pnw,Kyutoku:2017voj,Foucart:2019bxj,Metzger:2014ila,Siegel:2017nub,Siegel:2017jug,Fernandez:2018kax,Christie:2019lim,Fujibayashi:2020qda}, flat $Y_e$ distributions in $0.09$--$0.11$ and in $0.1$--$0.3$ are employed for the element abundances of the dynamical ejecta and the post-merger ejecta, respectively. We note that the recent study by numerical-relativity simulations for the remnant black hole-accretion torus system pointed out that the significant amount of post-merger ejecta with high values of $Y_e$ ($\gtrsim0.3$) may be driven even in the absence of the remnant massive neutron star if the ejection time scale is as long as $\gtrsim0.3\,{\rm s}$~\citep{Fujibayashi:2020qda}. As we show in Appendix~\ref{apx:YHcomp}, tighter upper limits to the ejecta mass are obtained for the post-merger ejecta with such high values of $Y_e$ than one with low values. Thus, the upper limits to the ejecta mass obtained in this work can be regarded as conservative limits. 

\cite{Wanajo:2014wha} pointed out that the spontaneous fissions of  $^{266}$Cf and $^{259,262}$Fm can significantly contribute to the heating rate particularly for $Y_e<0.1$. Since we employ the heating rate model of~\cite{Wanajo:2014wha}, these contributions are fiducially taken into account in our kilonova model. However, it is cautioned that the contribution of the spontaneous fissions to the heating rate is highly uncertain due to the uncertainty in the $\beta$-decay and spontaneous fission lifetimes of the parents nuclides~\citep[e.g.,][]{Wanajo:2014wha,Zhu:2018oay,Wanajo:2018wra}. The upper limit to the ejecta mass could depend on whether the contribution of the fission fragments to the heating rate is taken into account or not. Indeed, in our previous paper~\citep{Kawaguchi:2019nju}, we show that the fission fragments have significant contribution to enhancing the brightness of the kilonovae particularly for our BH-NS kilonova models. Thus, the radiative transfer calculation is performed also for the models without fission fragments to check how the upper limit could be affected by the uncertainty in the fission fragments. 

\section{Results}\label{sec:results}

\begin{figure*}
 	 \includegraphics[width=.33\linewidth]{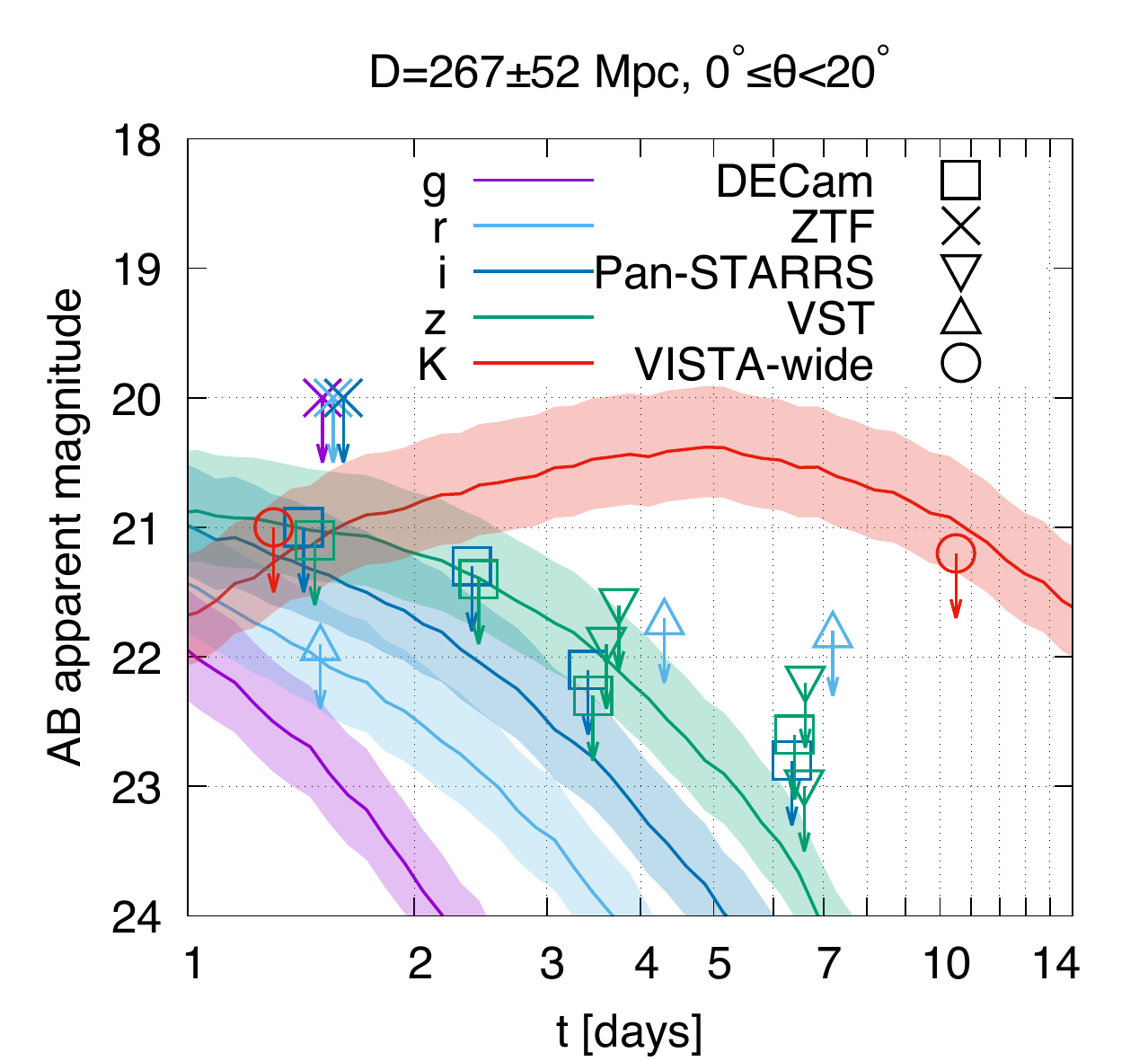}
 	 \includegraphics[width=.33\linewidth]{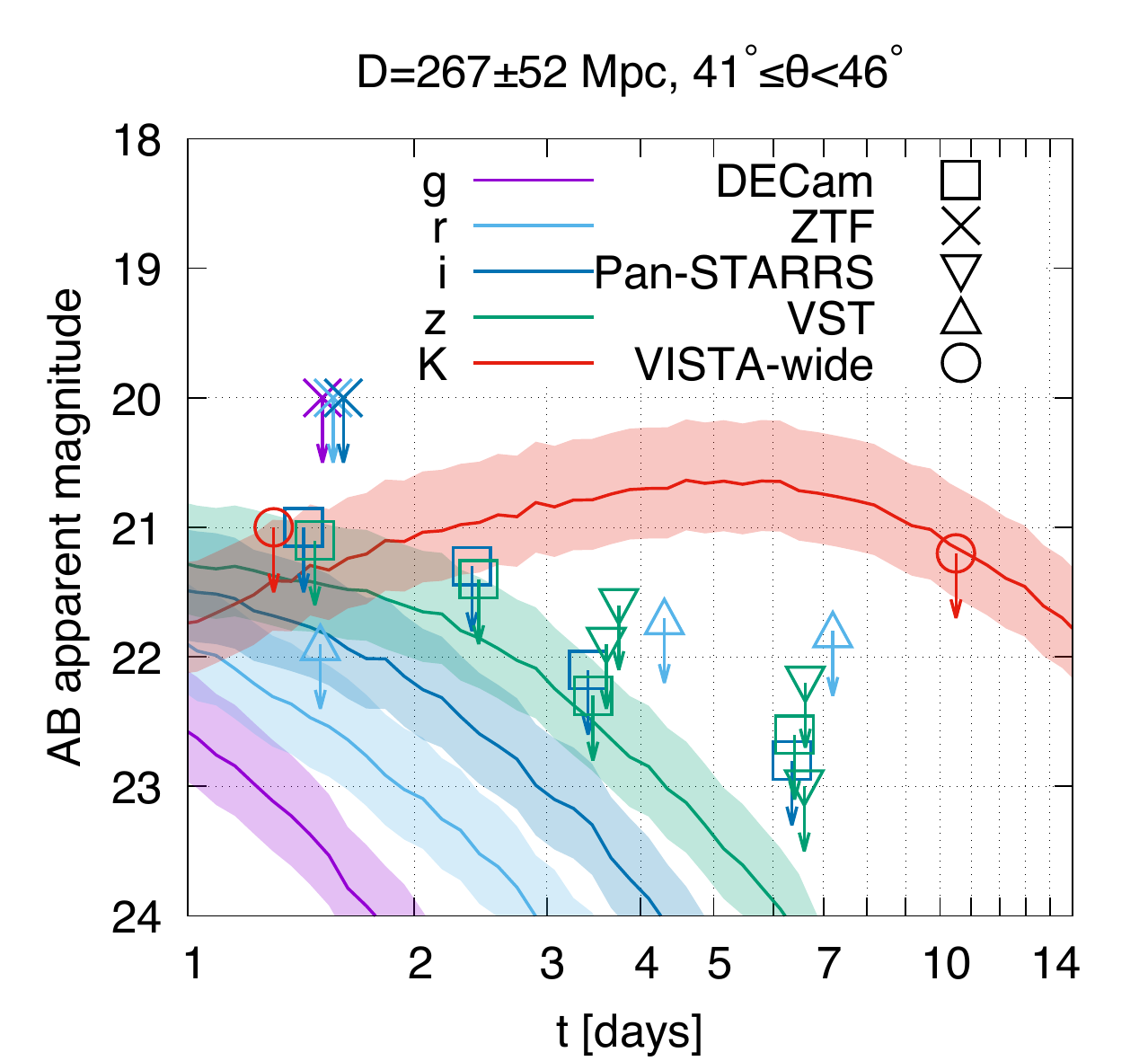}
 	 \includegraphics[width=.33\linewidth]{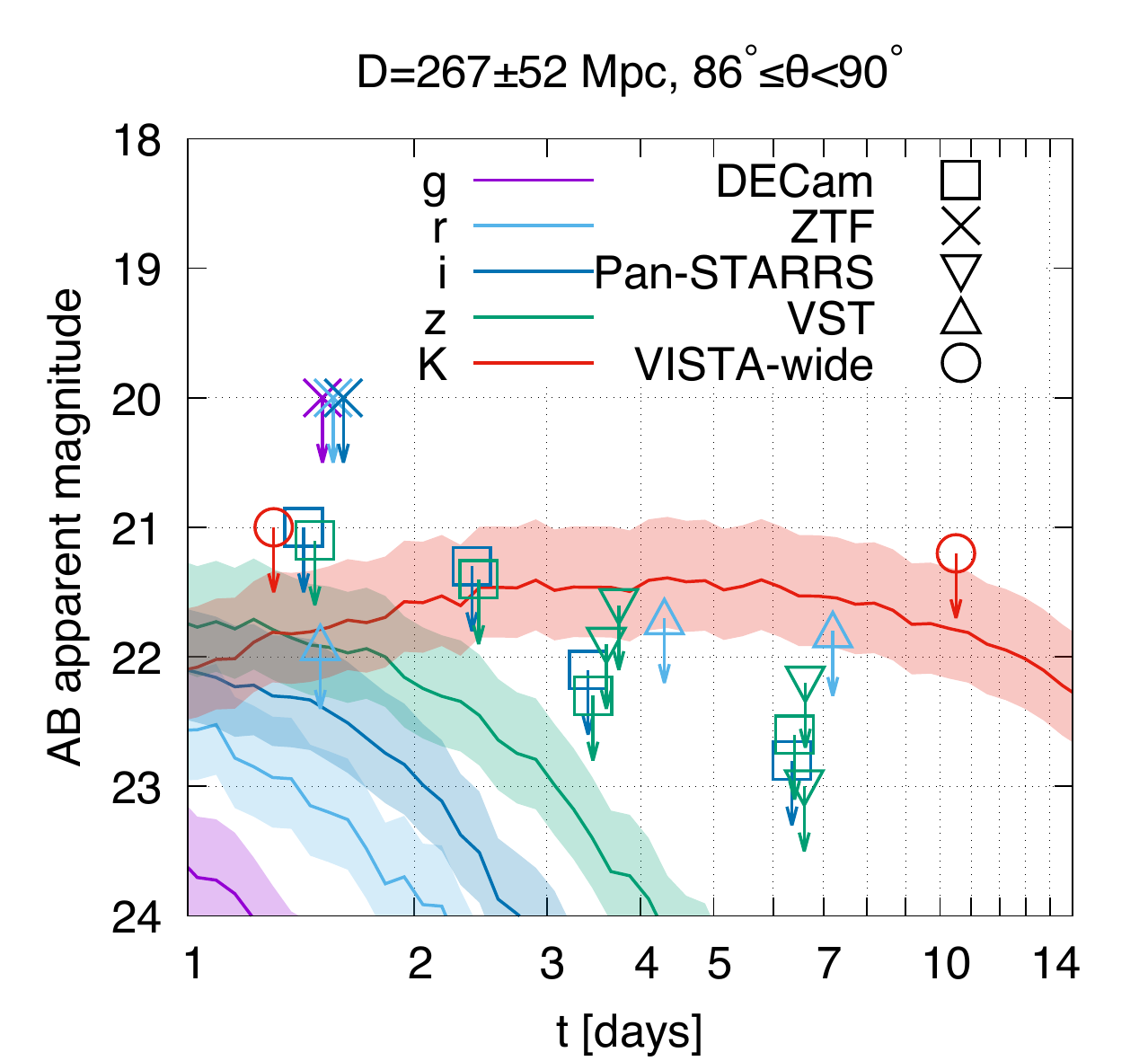}
 	 \caption{The {\it griz}-band light curves of a BH-NS kilonova model for $M_{\rm d}=0.02\,M_\odot$ and $M_{\rm pm}=0.02\,M_\odot$. The shaded regions denote the uncertainty in the brightness due to the error bar of the luminosity distance~\citep{gcn25333}. The upper limits to the EM counterparts obtained by DECam~\citep{Andreoni:2019qgh}, ZTF~\citep{gcn25381}, Pan-STARRS, VST, and VISTA~\citep{Ackley:2020qkz} covering $70$--$97$\% of the sky localization probability~\citep{gcn25333} are shown. The purple, light blue, blue, green, and red curves and points denote the lightcurves and upper limits for the {\it g}, {\it r}, {\it i}, {\it z}, and {\it K} band filters, respectively.}
	 \label{fig:mw002_md002}
\end{figure*}

\begin{figure*}
 	 \includegraphics[width=.33\linewidth]{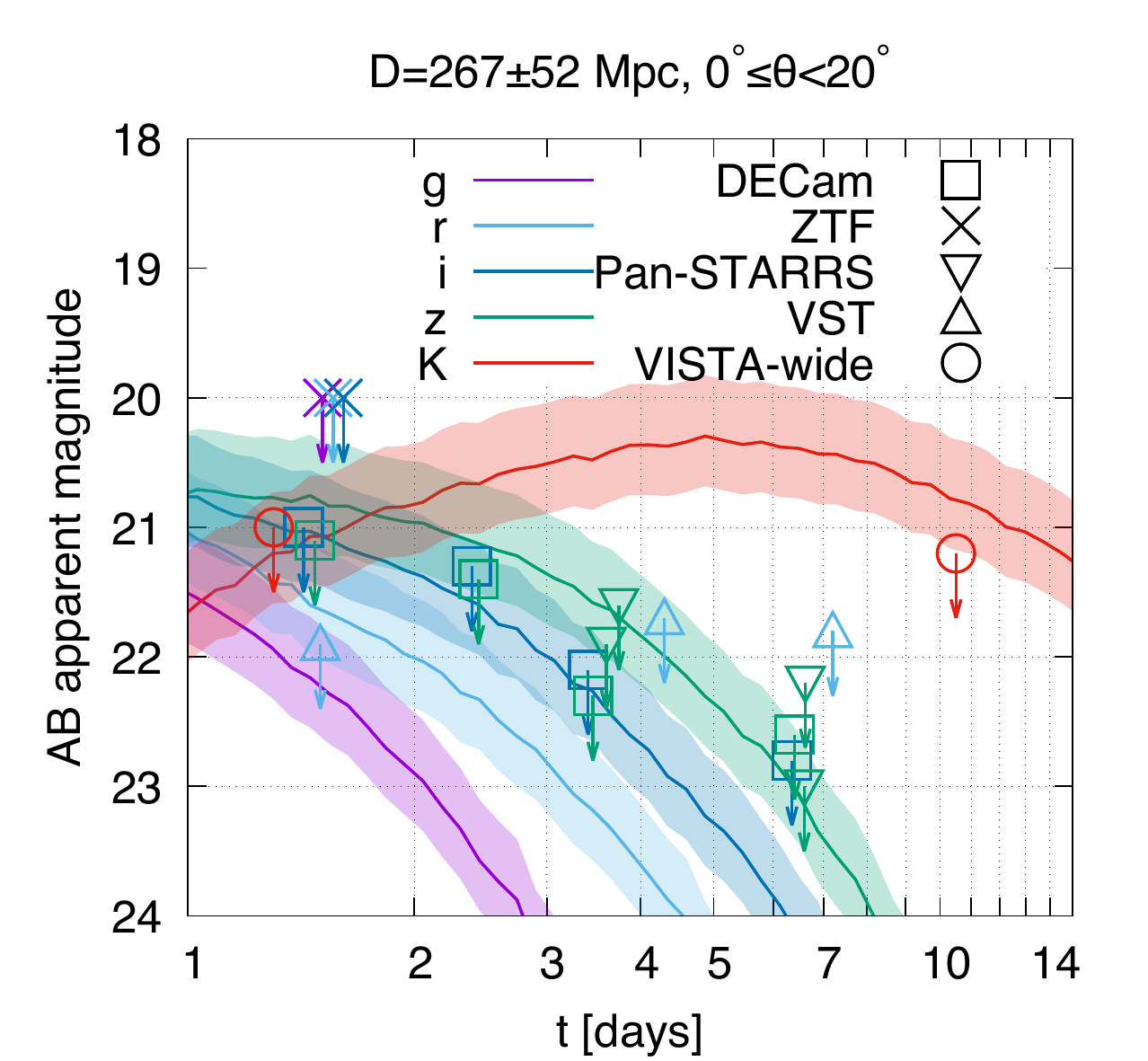}
 	 \includegraphics[width=.33\linewidth]{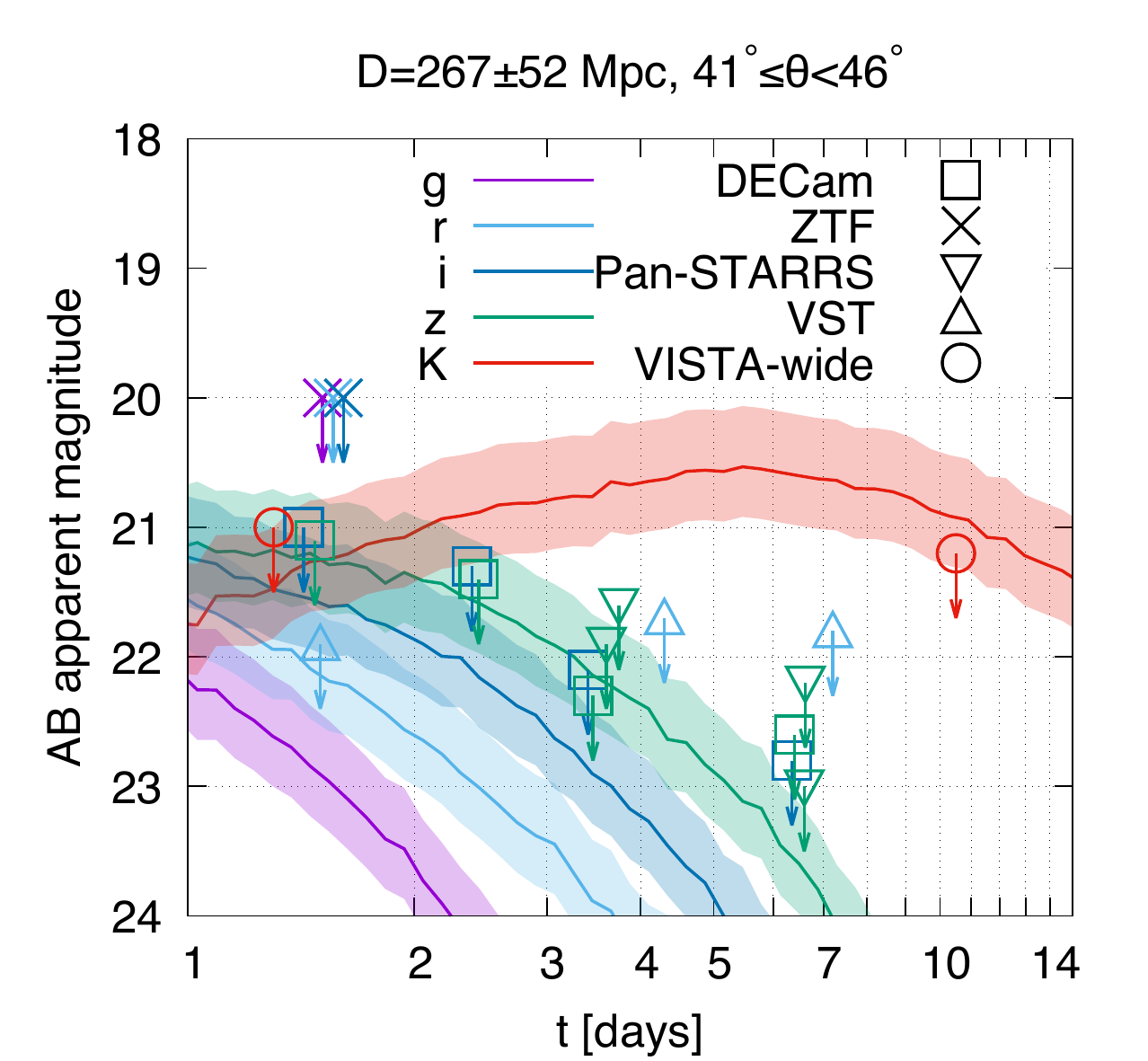}
 	 \includegraphics[width=.33\linewidth]{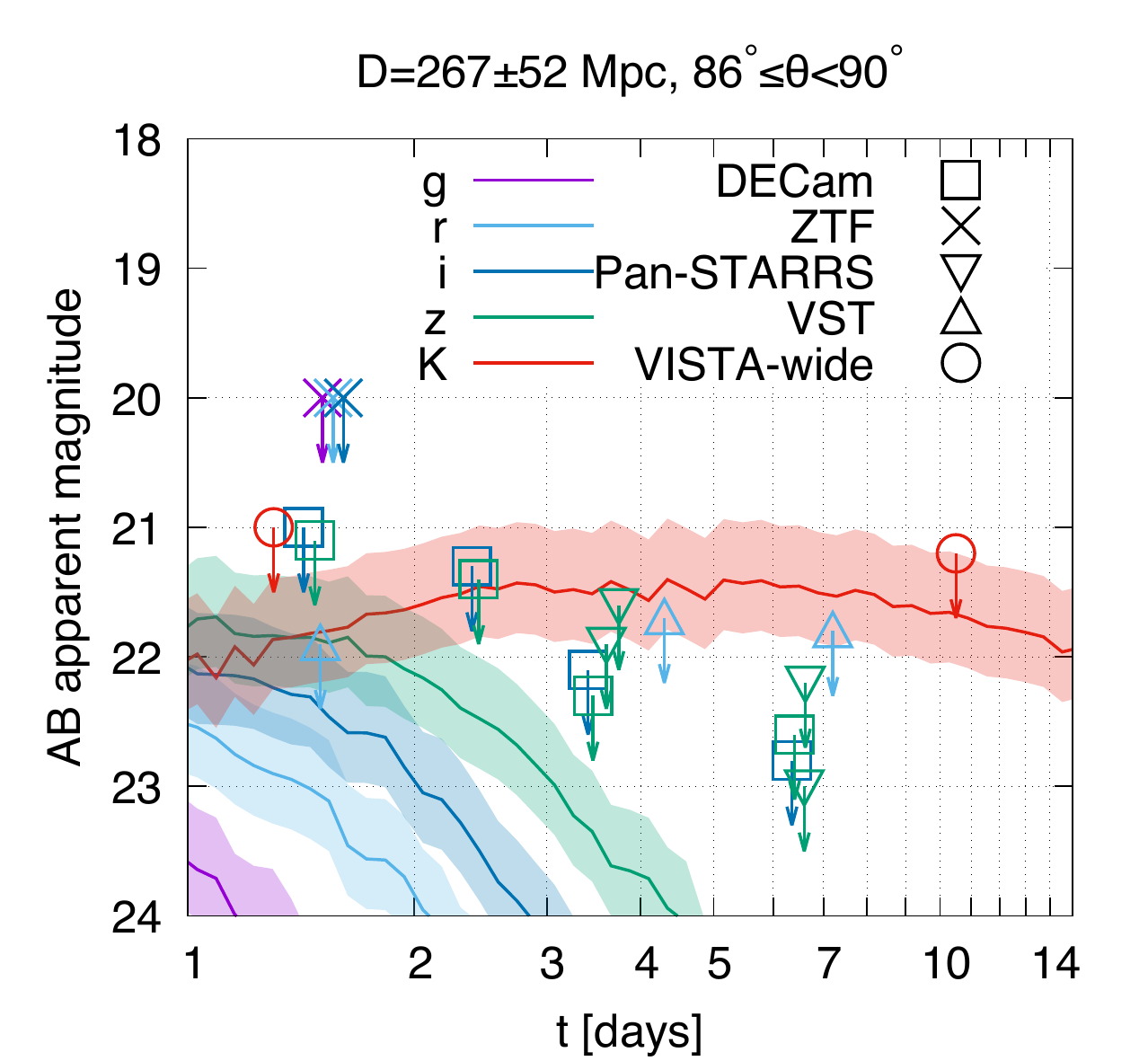}
 	 \caption{The same as Figure~\ref{fig:mw002_md002} but for the model with $M_{\rm d}=0.02\,M_\odot$ and $M_{\rm pm}=0.04\,M_\odot$.}
	 \label{fig:mw004_md002}
\end{figure*}

Figure~\ref{fig:mw002_md002} shows the {\it griz}-band lightcurves observed from $0^\circ\le\theta_{\rm obs}\le20^\circ$, $41^\circ\le\theta_{\rm obs}\le46^\circ$, and $86^\circ\le\theta_{\rm obs}\le90^\circ$ for the model with $M_{\rm d}=0.02\,M_\odot$ and $M_{\rm pm}=0.02\,M_\odot$, where $\theta_{\rm obs}$ denotes the angle of the observer measured from the axis of symmetry. In Figure~\ref{fig:mw002_md002}, we also show the upper limits to the EM counterparts obtained by DECam~\citep{Andreoni:2019qgh}, ZTF~\citep{gcn25381}, Pan-STARRS, VST, and VISTA~\citep{Ackley:2020qkz} covering $70$--$97$\% of the sky localization probability~\citep{gcn25333}. Here, $t$ denotes the day since the merger time. Figure~\ref{fig:mw004_md002} is the same as Figure~\ref{fig:mw002_md002} but for the model with $M_{\rm d}=0.02\,M_\odot$ and $M_{\rm pm}=0.04\,M_\odot$. As Figures~\ref{fig:mw002_md002} and \ref{fig:mw004_md002} indicate, for the models with the same amount of the dynamical ejecta, the brightness in the {\it griz}-bands increases as the post-merger ejecta mass increases. 
 
The brightness of the kilonova model depends on the viewing angle in the presence of dynamical ejecta reflecting its non-spherical density profile. As $\theta_{\rm obs}$ increases, the emission in the {\it griz}-band becomes faint, and hence, the upper limit to the ejecta mass becomes weaker approximately monotonically (see also Figure~\ref{fig:magrdw}). This is due to blocking of photons emitted from the post-merger ejecta by the dynamical ejecta~\citep{Kasen:2014toa,Kawaguchi:2018ptg,Bulla:2019muo,Kawaguchi:2019nju}, which is enhanced as $\theta_{\rm obs}$ increases since the density of dynamical ejecta is high around the equatorial plane. 

We find that the upper limit to the {\it z}-band at 3.43 d after the GW trigger obtained by DECam~\citep{Andreoni:2019qgh} and the upper limit to the {\it K}-band at 9.2--10.5 d by VISTA~\citep{Ackley:2020qkz} provide the tightest constraint on the kilonova lightcurves, which cover $97\%$ and $94\%$ of the sky localization probability~\citep{gcn25333}, respectively. Indeed, we find that the other upper limits are always satisfied as far as the kilonova model is consistent with these upper limits. Thus, in the following, we focus on these upper limits to constrain the ejecta mass. Note that, to obtain conservative results, we adapt 10.5 d as the time of the upper limit to the {\it K}-band obtained by VISTA~\citep{Ackley:2020qkz}, since the {\it K}-band brightness is decreasing in such phase for the kilonova models studied in this work.

\subsection{The upper limit to the total ejecta mass}
In this subsection, we focus on the upper limit to the total ejecta mass. To derive a conservative result, we first explore how the faintest emission is obtained among the fixed total ejecta mass models. Then, we argue the upper limit to the total ejecta mass based on the models.

\begin{figure*}
 	 \includegraphics[width=.48\linewidth]{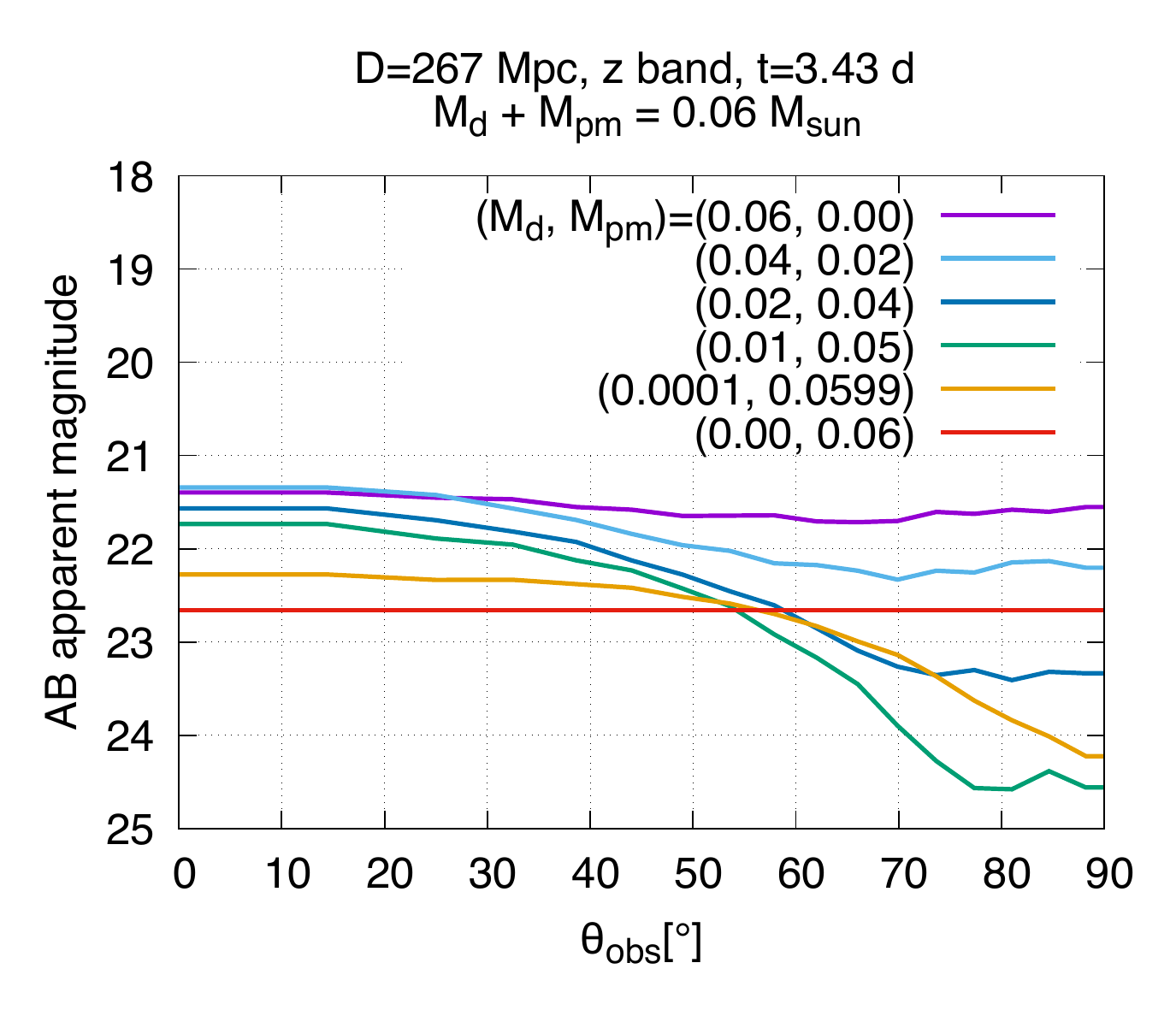}
 	 \includegraphics[width=.48\linewidth]{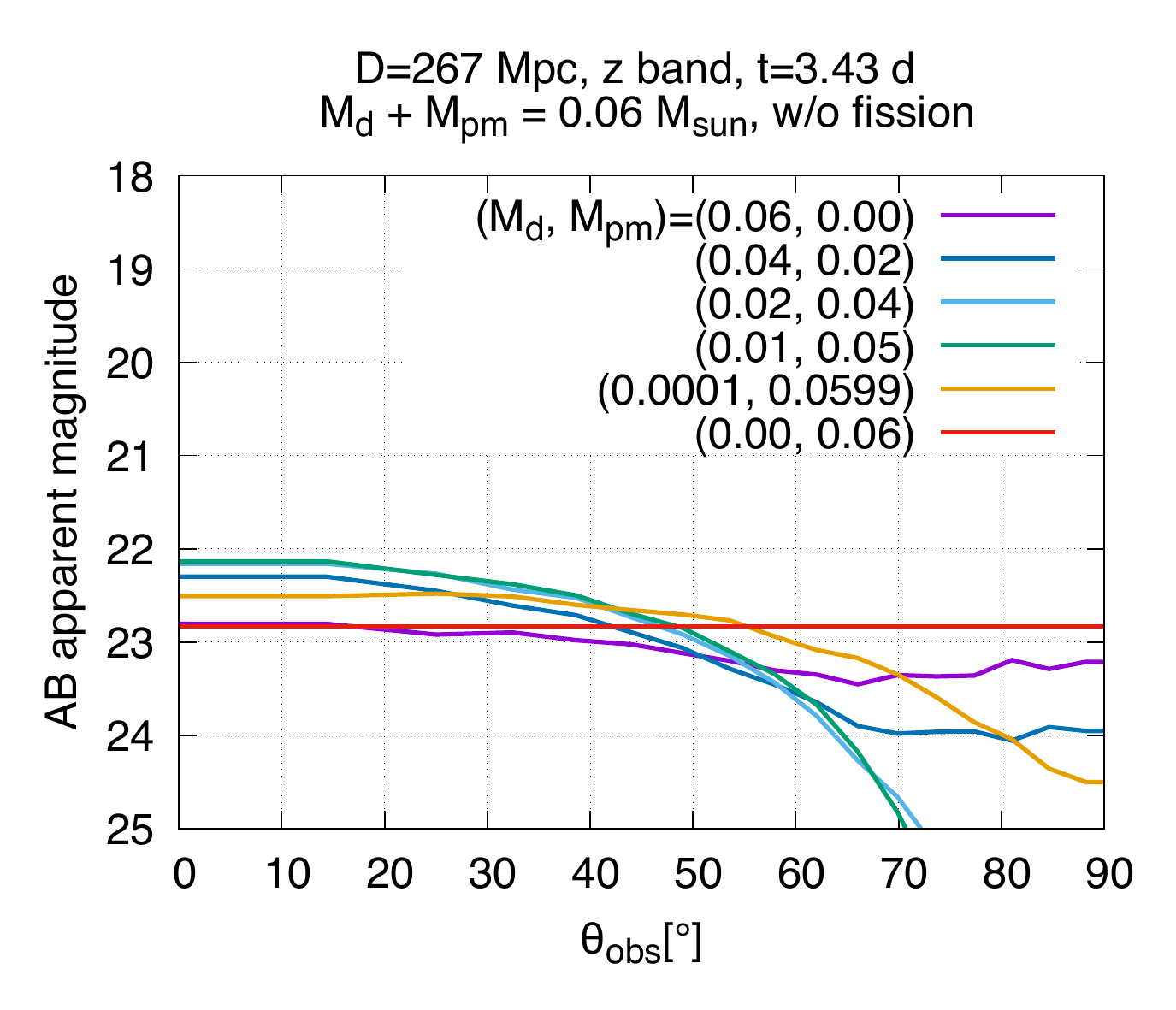}
 	 \caption{The brightness of the {\it z}-band emission at $t=3.43\,{\rm d}$ as a function of the viewing angle, $\theta_{\rm obs}$, for several models with $M_{\rm d}+M_{\rm pm}=0.06\,M_\odot$. The values in the legend denote $(M_{\rm d},M_{\rm pm})$ in the unit of $M_\odot$. The left and right panels show the results that take into account and omit the contribution from the fission fragments to the heating rate, respectively. We note that there is no viewing angle dependence for the model only with the spherical post-merger ejecta.}
	 \label{fig:magrdw}
\end{figure*}
\begin{figure*}
 	 \includegraphics[width=.48\linewidth]{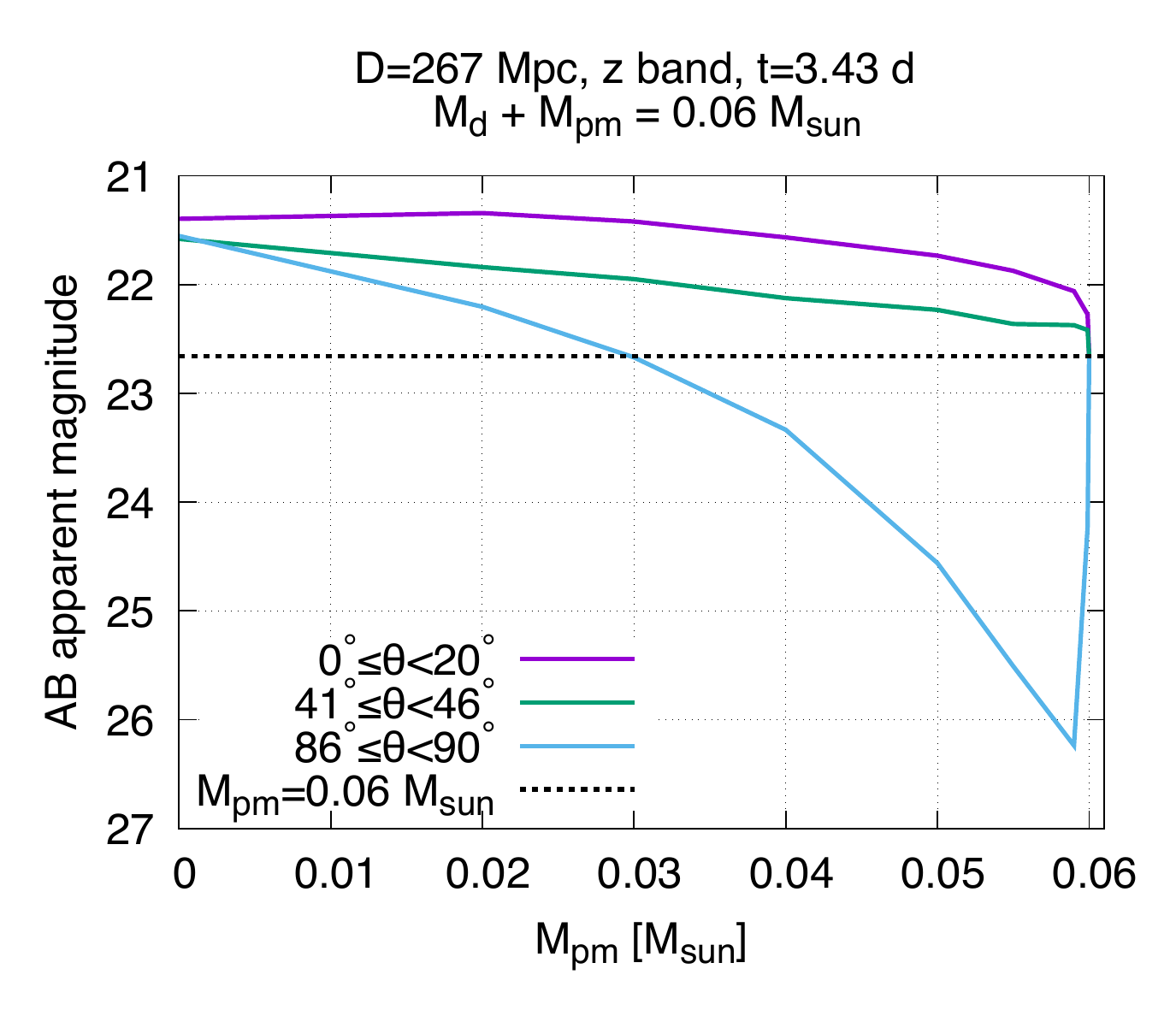}
 	 \includegraphics[width=.48\linewidth]{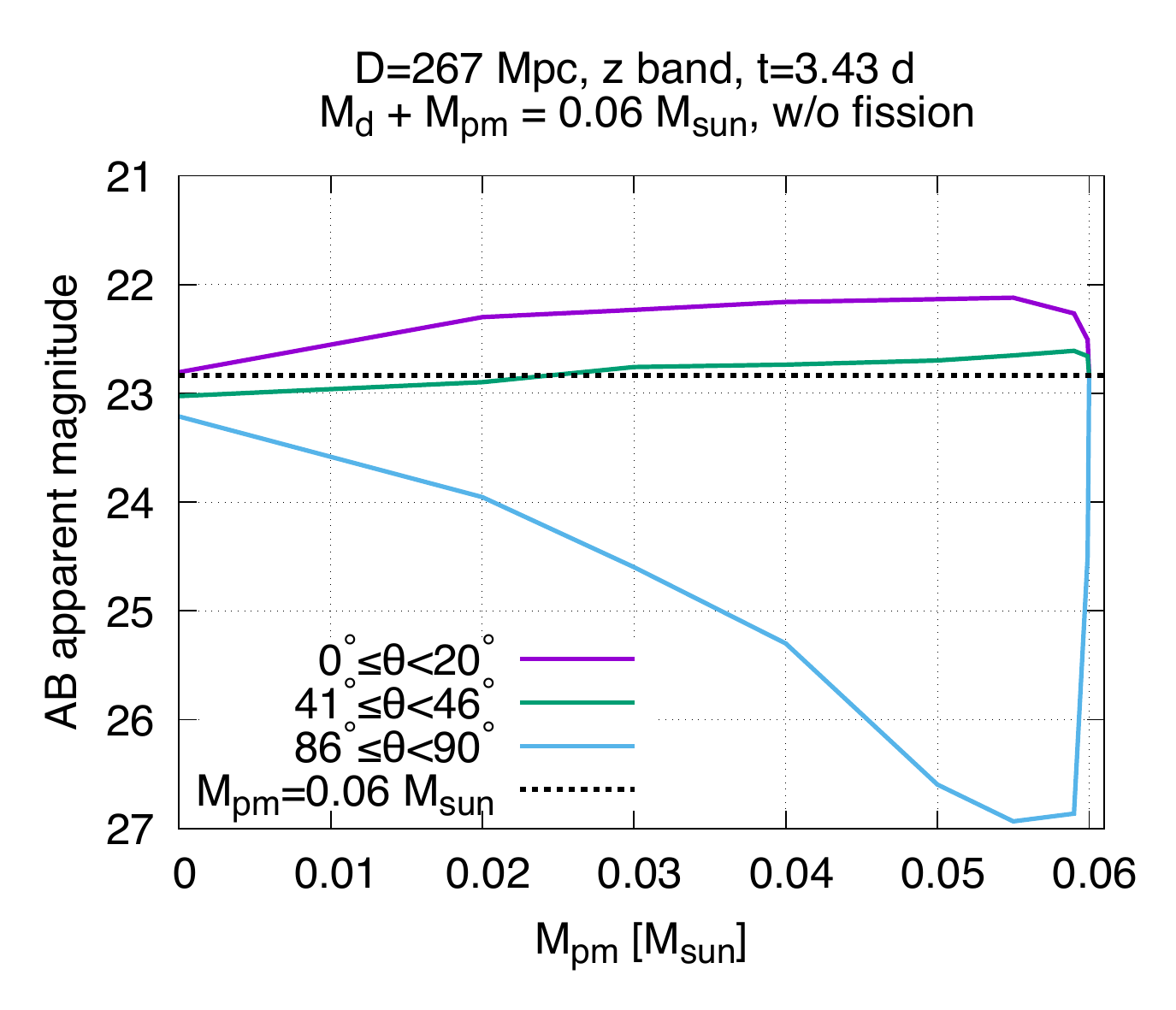}
 	 \caption{The brightness of the {\it z}-band emission at $t=3.43\,{\rm d}$ as a function of $M_{\rm pm}$ for the models with $M_{\rm d}+M_{\rm pm}=0.06\,M_\odot$. The left and right panels show the results that take into account and omit the contribution from the fission fragments to the heating rate, respectively. The dashed line denotes the brightness for the model with $M_{\rm pm}=0.06\,M_\odot$ and $M_{\rm d}=0.00\,M_\odot$.}
	 \label{fig:magmpm}
\end{figure*}

The brightness of the emission depends on the ratio between the dynamical and post-merger ejecta mass among the fixed total ejecta mass models. As an illustration, Figure~\ref{fig:magrdw} and Figure~\ref{fig:magmpm} show the brightness of the {\it z}-band emission at $t=3.43\,{\rm d}$ as functions of $\theta_{\rm obs}$ and $M_{\rm pm}$ for the models with $M_{\rm d}+M_{\rm pm}=0.06\,M_\odot$, respectively. For $\theta_{\rm obs}\lesssim45^\circ$, we find that the {\it z}-band brightness at $t=3.43\,{\rm d}$ for the models with the same total ejecta mass increases approximately monotonically as the ratio $M_{\rm d}/M_{\rm pm}$ increases, and the faintest emission is realized for the model only with the post-merger ejecta if the fission fragment is taken into account. This is mostly due to the fact that the specific deposition rate of thermal photons, which is determined by the radioactive heating rate and the thermalization efficiency, is higher for the dynamical ejecta than the post-merger ejecta. We note that the difference in the opacity is also responsible for the difference in the brightness. However, for $\theta_{\rm obs} \alt 45^\circ$, this effect is minor. 

In contrast, the dependence of the {\it z}-band brightness at $t=3.43\,{\rm d}$ on the ratio $M_{\rm d}/M_{\rm pm}$ is more complicated for $\theta_{\rm obs}\gtrsim45^\circ$. The emission becomes faint as the ratio $M_{\rm d}/M_{\rm pm}$ decreases for $M_{\rm d}\gtrsim0.01\,M_\odot$, but the brightness increases again for $M_{\rm d}\lesssim0.001\,M_\odot$. This is due to the fact that, with the decrease of $M_{\rm d}$, the emission from the dynamical ejecta becomes less significant and only its own blocking effect of photons plays a role~\citep{Kasen:2014toa,Kawaguchi:2018ptg,Bulla:2019muo,Kawaguchi:2019nju}. For such a situation, the emission becomes bright as the dynamical ejecta mass decreases.

Dependence of the emission on the ratio $M_{\rm d}/M_{\rm pm}$ is different for the case in which the contribution from the fission fragments to the heating rate is omitted. For such cases, the {\it z}-band brightness observed from $\theta_{\rm obs}\lesssim45^\circ$ at $t=3.43\,{\rm d}$ becomes the brightest for $M_{\rm d}/M_{\rm pm}\approx20$--$50\%$, and it becomes faint as the ratio increases. Nevertheless, the faintest emission is realized approximately for the model only with the post-merger ejecta for $\theta_{\rm obs}\lesssim45^\circ$. The dependence of the {\it z}-band brightness on the ratio $M_{\rm d}/M_{\rm pm}$ for $\theta_{\rm obs}\gtrsim45^\circ$ is qualitatively similar to what is found for the models with the fission fragments. However, the emission is fainter by 1.5 mag than those with the fission fragments, and the model only with the post-merger ejecta always gives the brightest lightcurve.

 We find that the brightness of the {\it K}-band emission at $t=10.5\,{\rm d}$ also shows broadly the same dependence on the ratio $M_{\rm d}/M_{\rm pm}$ as for the {\it z}-band emission at $t=3.43\,{\rm d}$. For the {\it K}-band emission at $t=10.5\,{\rm d}$, the decrease of the brightness due to the blocking effect is much less significant, and the difference in the emission observed from the polar and equatorial directions are $\lesssim 1.5\,{\rm mag}$.

\begin{figure}
 	 \includegraphics[width=1\linewidth]{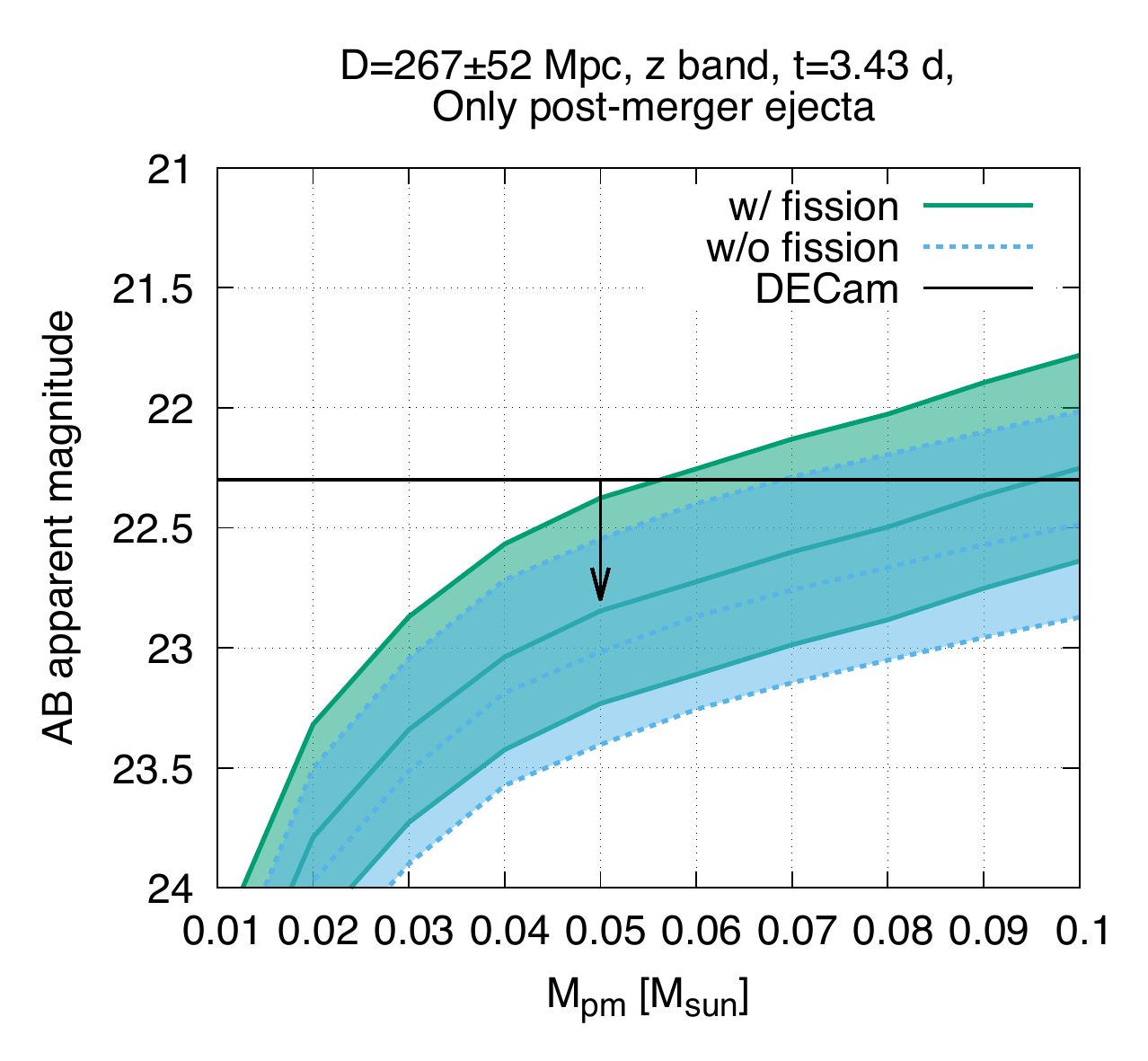}
 	 \caption{The brightness of the {\it z}-band emission at $t=3.43\,{\rm d}$ as a function of ejecta mass for the model only with the post-merger ejecta. The curves plotted at the center of filled regions denote the brightness for $D=267$ Mpc, while the curves plotted at the lower and upper edges denote the brightness assuming $D=267\pm52$ Mpc, respectively~\citep{gcn25333}. The black horizontal line shows the upper limit to the {\it z}-band emission at 3.43 d for S190814bv obtained by DECam~\citep{Andreoni:2019qgh}. We note that there is no viewing angle dependence for the model only with the spherical post-merger ejecta.}
	 \label{fig:magpm}
\end{figure}

If $\theta_{\rm obs} \leq 45^\circ$, the model only with the post-merger ejecta provides the conservative upper limit to the total ejecta mass. We find that the upper limit to the {\it z}-band emission at 3.43 d provides the tightest constraint for this setup. Figure~\ref{fig:magpm} shows the brightness of the {\it z}-band emission at $t=3.43\,{\rm d}$ as a function of ejecta mass for the model only with the post-merger ejecta. For $D \geq 267$\,Mpc, only weak upper limits are obtained, and the model with $\lesssim0.1\,M_\odot$ is always consistent with the upper limit to the emission. On the other hand, assuming an optimistic distance of $215$ Mpc, the upper limit to the emission obtained by~\cite{Andreoni:2019qgh} implies that the total ejecta mass should be less than $\approx0.06\,M_\odot$ and $\approx0.07\,M_\odot$ for the models with and without the fission fragments, respectively.

The constrains become much weaker if $\theta_{\rm obs}\gtrsim60^\circ$, because the faintest emission for this case is realized in the presence of a fraction of the dynamical ejecta; for the case that the blocking effect of the dynamical ejecta is significant. The {\it z}-band emission is suppressed by 1--2 mag than the case with the same total ejecta mass but only with the post-merger ejecta. For a large viewing angle ($\theta_{\rm obs}\gtrsim60^\circ$), the upper limit to the {\it K}-band emission provides the tightest constraint, since the suppression due to the blocking effect is less significant for the $K$-band emission. Nevertheless, we find that the total ejecta mass as large as $0.1\,M_\odot$ is consistent with the observation even assuming $D=215$ Mpc.

\subsection{The upper limit to the dynamical ejecta mass}\label{sec:results_dyn}
In this section, we focus on the upper limit to the dynamical ejecta mass. As shown in the previous subsection, the {\it z}-band brightness observed from $\theta_{\rm obs}\lesssim45^\circ$ at $t=3.43\,{\rm d}$ increases as the ratio $M_{\rm d}/M_{\rm pm}$ increases for the fixed total ejecta mass and for the case that the fission fragments plays an important role. This indicates that the obtained upper limit to the emission could be informative to constrain the dynamical ejecta mass. Furthermore, for BH-NS mergers, the connection between the dynamical ejecta mass and the binary parameters, such as the mass of each component, BH spin, and NS radius, is expected to be predicted relatively more accurately by numerical-relativity simulations than the post-merger ejecta or the dynamical ejecta for NS-NS mergers~\citep{Hotokezaka:2012ze,Kyutoku:2015gda,Dietrich:2016fpt,Foucart:2018rjc}. This is because the dynamical ejecta is driven approximately purely by gravitational torque for BH-NS mergers, while shocks and magnetically driven turbulence also play important roles for the others. Thus, the constraint on the dynamical ejecta mass could be useful for constraining parameters of observed binaries as we discuss in Section~\ref{sec:discussion}.

Numerical-relativity simulations for BH-NS mergers suggest that remnant torus (gravitationally bounded component of the material that remains after the merger) is typically more massive than the dynamical ejecta~\citep[e.g.,][]{Kyutoku:2015gda,Foucart:2019bxj}. This indicates that a significant amount of the post-merger ejecta would always be accompanied with the massive dynamical ejecta. In fact, for example, numerical-relativity simulations for BH-NS mergers~\citep[e.g.,][]{Kyutoku:2015gda,Foucart:2019bxj} show that the mass of the remnant torus is typically larger than the dynamical ejecta mass by a factor of more than 3. Magneto-hydrodynamics or viscous hydrodynamics simulations for the BH-accretion torus systems~\citep{Fernandez:2013tya,Metzger:2014ila,Just:2014fka,Siegel:2017nub,Siegel:2017jug,Fernandez:2018kax,Christie:2019lim,Fujibayashi:2020qda} suggest that $\approx20$--$30\%$ of the remnant torus could be ejected from the system. Hence, the post-merger ejecta mass is likely to be larger than or comparable with the dynamical ejecta mass. In the following, we focus particularly on the models with $M_{\rm pm} = M_{\rm d}$. This is because the brightness in the {\it griz}-bands for a given epoch $\gtrsim 0.3$ d increases monotonically as the post-merger ejecta mass increases for the models with a fixed amount of the dynamical ejecta, and thus, the models with $M_{\rm pm} = M_{\rm d}$ provide a conservative upper limit to the dynamical ejecta mass for given upper limit to the emission as long as focusing on the cases of $M_{\rm pm}\ge M_{\rm d}$ (see Appendix~\ref{apx:mpm05md} for the upper limit to the dynamical ejecta mass assuming more conservative setup, $M_{\rm pm}=0.5M_{\rm d}$).

\begin{figure*}
 	 \includegraphics[width=.5\linewidth]{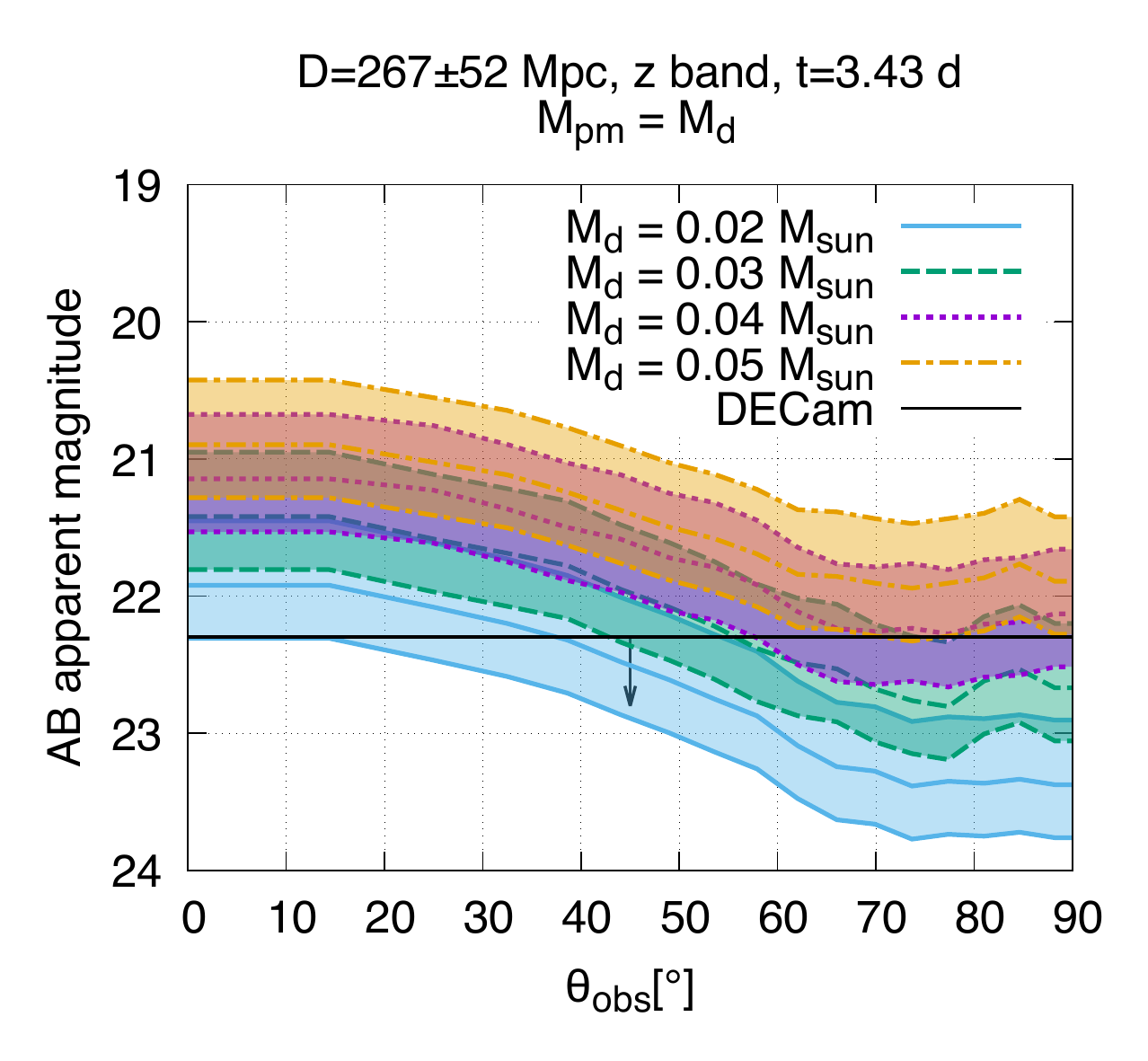}
 	 \includegraphics[width=.5\linewidth]{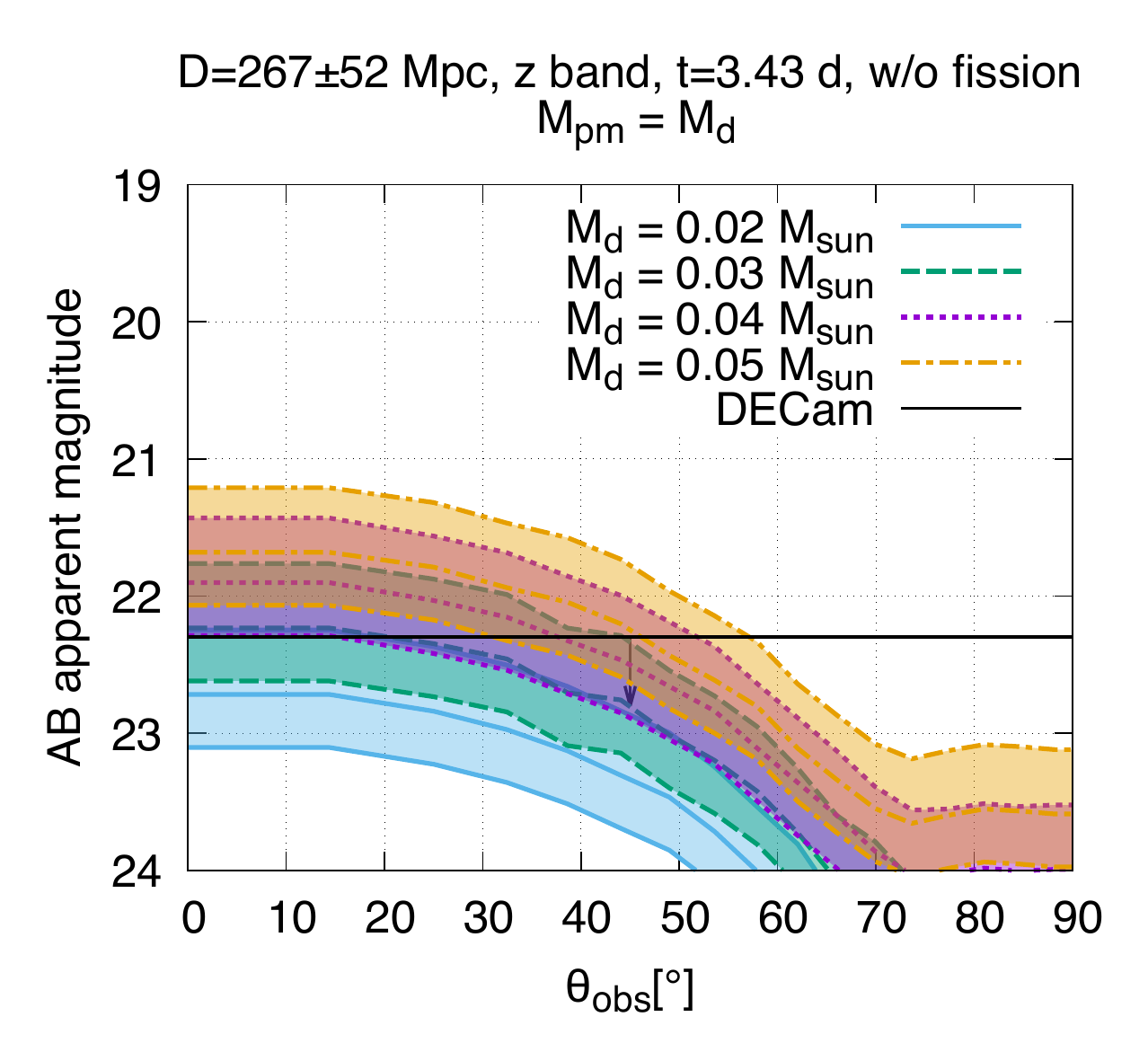}
 	 \caption{The brightness of the {\it z}-band emission at $t=3.43\,{\rm d}$ as a function of viewing angle, $\theta_{\rm obs}$. The left panel shows the lightcurves for the models with $(M_{\rm d},M_{\rm pm})=(0.02\,M_\odot, 0.02\,M_\odot)$ (blue solid), $(0.03\,M_\odot, 0.03\,M_\odot)$ (green dashed), $(0.04\,M_\odot, 0.04\,M_\odot)$ (purple dotted), and $(0.05\,M_\odot, 0.05\,M_\odot)$ (orange dotted). The right panel is the same as the left panel but for the models in which the contribution from the fission fragments to the heating rate is omitted. The curves plotted at the center of shaded regions denote the brightness for $D=267$ Mpc, while the curves plotted at the lower and upper edges denote the brightness assuming $D=267\pm52$ Mpc, respectively~\citep{gcn25333}. The black horizontal lines in the left and right plots show the upper limits to the {\it z}-band emission at 3.43 d for S190814bv obtained by DECam~\citep{Andreoni:2019qgh}.}
	 \label{fig:mag_comp}
\end{figure*}

\begin{figure*}
 	 \includegraphics[width=.5\linewidth]{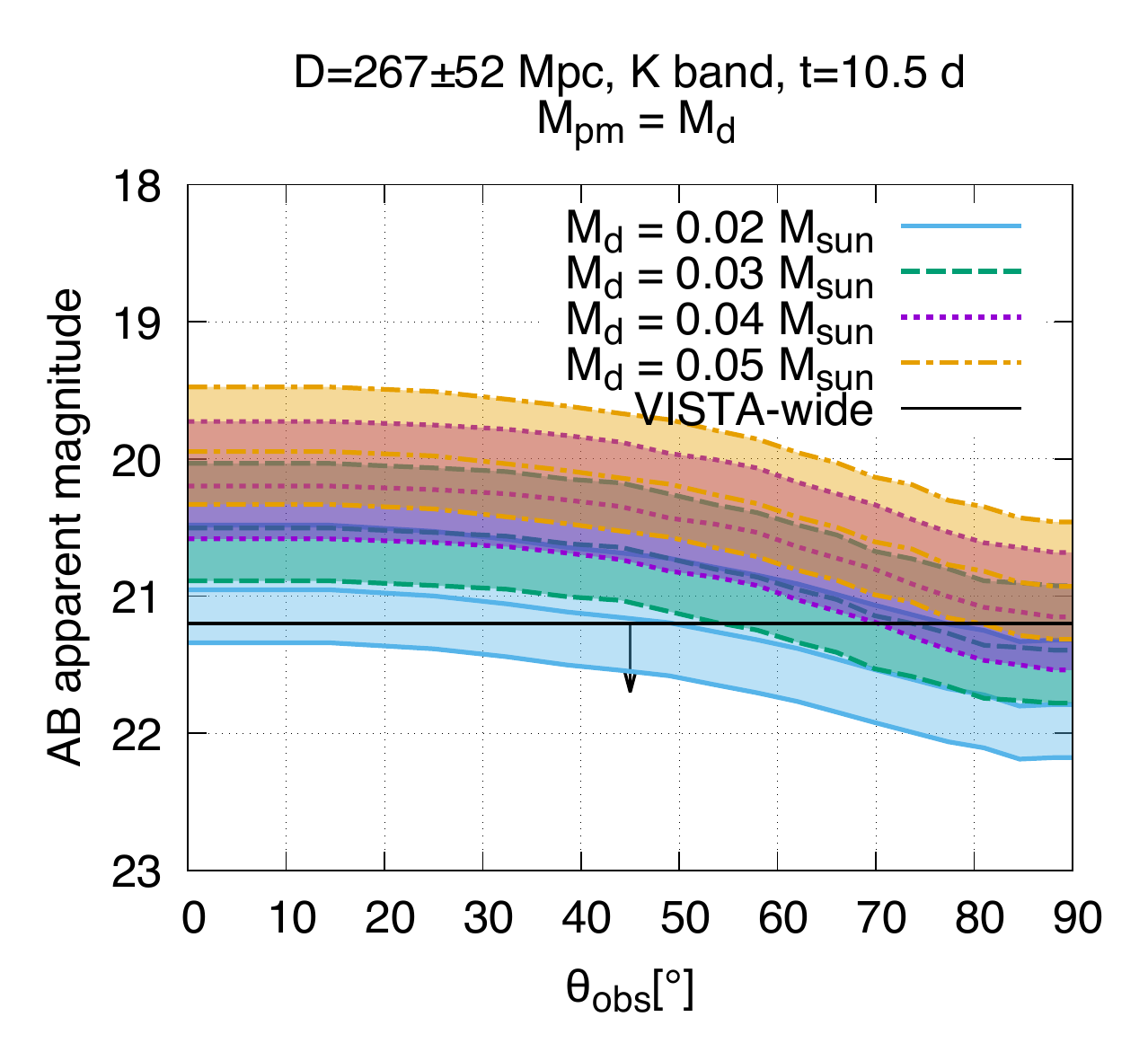}
 	 \includegraphics[width=.5\linewidth]{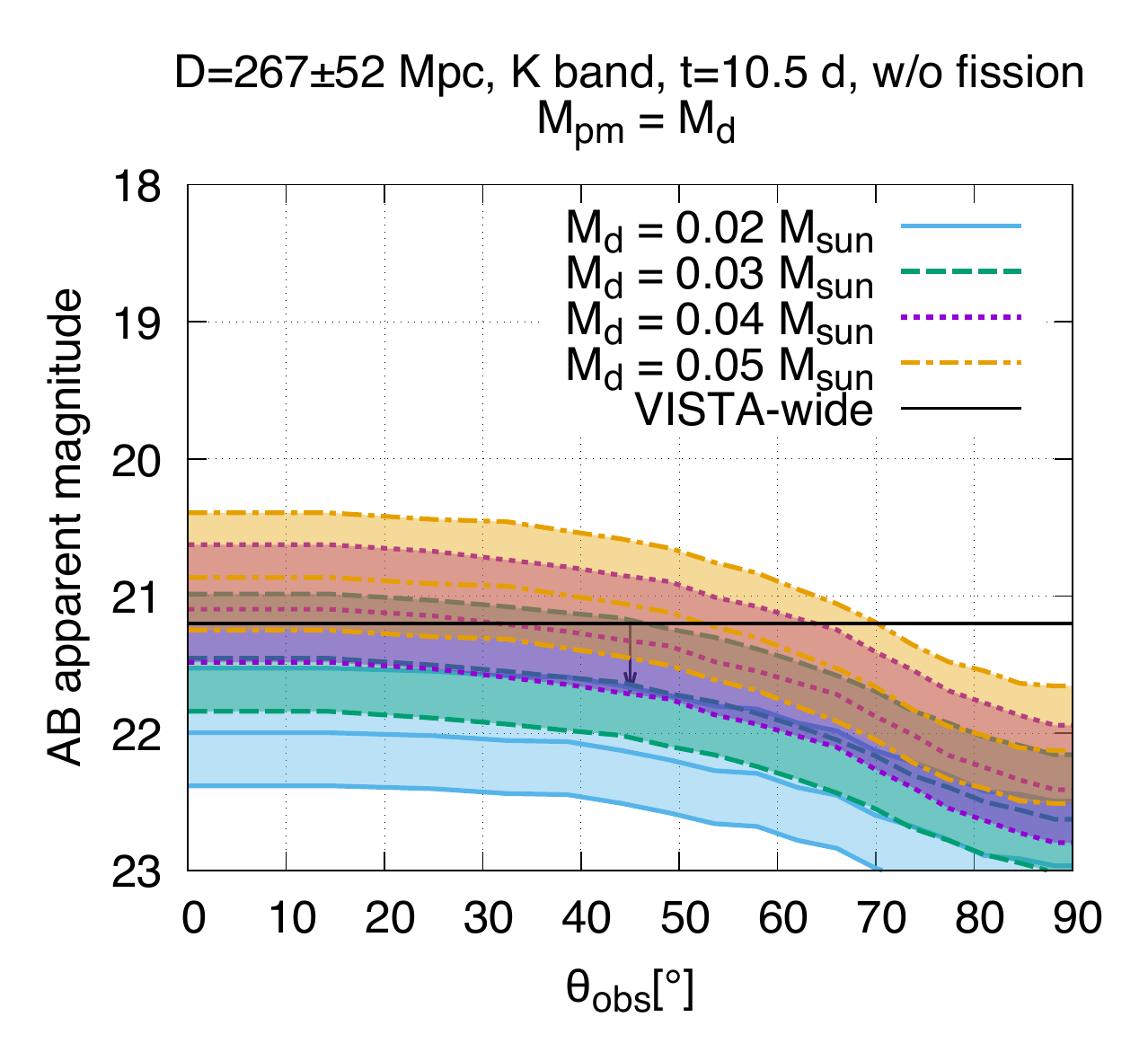}
 	 \caption{The same as Figure~\ref{fig:mag_comp} but for the brightness of the {\it K}-band emission at $t=10.5\,{\rm d}$ with the upper limit obtained by VISTA~\citep{Ackley:2020qkz}.}
	 \label{fig:mag_comp_K}
\end{figure*}

The left panel in Figure~\ref{fig:mag_comp} shows the {\it z}-band brightness at 3.43 d after the merger as a function of $\theta_{\rm obs}$ for the models with $(M_{\rm d},M_{\rm pm})=(0.02\,M_\odot, 0.02\,M_\odot)$, $(0.03\,M_\odot, 0.03\,M_\odot)$, $(0.04\,M_\odot, 0.04\,M_\odot)$, and $(0.05\,M_\odot, 0.05\,M_\odot)$ together with the upper limit by DECam~\citep{Andreoni:2019qgh}. In the following, we focus on the results assuming $D=319$ Mpc to obtain conservative upper limits. Here, $D=319$ Mpc is the $1\sigma$ far edge of the posterior inferred by the GW analysis~\citep{gcn25333}, and the lower edge of the shaded region in Figure~\ref{fig:mag_comp} corresponds to the predicted brightness for $D=319$ Mpc. We note that the estimated distance should depend on the viewing angle,  and larger and smaller distance would be favored for the face-on and edge-on observation, respectively, due to the correlation in determining GW amplitude (for example, see ~\cite{Abbott:2020uma} for the case of GW190425).

If $\theta_{\rm obs}\leq45^\circ$, the emission in the {\it z}-band at 3.43 d for the model with $(M_{\rm d},M_{\rm pm})=(0.03\,M_\odot, 0.03\,M_\odot)$ is brighter than 22.3 mag for the inferred $1\,\sigma$ range of the luminosity distance. This indicates that the ejecta with $M_{\rm d}=M_{\rm pm}\geq0.03\,M_\odot$ is unlikely to be driven in S190814bv if $\theta_{\rm obs}\le45^\circ$. For a smaller value of $\theta_{\rm obs}$, the upper limit to the ejecta mass becomes tighter. For $\theta_{\rm obs}\le20^\circ$, the model with $(M_{\rm d},M_{\rm pm})=(0.02\,M_\odot, 0.02\,M_\odot)$ is disfavored or only marginally consistent with the upper limit to the {\it z}-band emission at 3.43 d. On the other hand, the models with $M_{\rm pm}=M_{\rm d} \leq0.04\,M_\odot$ cannot be ruled out if $\theta_{\rm obs}\gtrsim60^\circ$. The models with $M_{\rm pm}=M_{\rm d} \geq0.05\,M_\odot$ is always disfavored regardless of the viewing angle.

The right panel of Figure~\ref{fig:mag_comp} shows the same as the left panel but for the models in which the contribution from the fission fragments to the heating rate is omitted. The {\it z}-band emission becomes fainter by $\approx1$ and $\approx2$ mag for the polar and equatorial direction, respectively, than the results shown in the left panel of Figure~\ref{fig:mag_comp}. The brightness observed from the equatorial direction is affected more significantly than that observed from the polar direction by omitting the fission fragments because it is dominated by the emission from the dynamical ejecta in our models, in which the fission fragments have a large impact on the heating rate. Due to the fainter emission, the upper limit to the ejecta mass is weaker for the models without the fission fragments. If $\theta_{\rm obs}$ is larger than $30^\circ$, the model with $M_{\rm d}\geq0.05\,M_\odot$ is consistent with the upper limit to the emission. On the other hand, if $\theta_{\rm obs}$ is smaller than $30^\circ$, the models only with $M_{\rm d} \leq0.04\,M_\odot$ is allowed for the assumption with $M_{\rm pm}\ge M_{\rm d}$.

Figure~\ref{fig:mag_comp_K} shows the same as Figure~\ref{fig:mag_comp} but for the brightness of the {\it K}-band emission at $t=10.5\,{\rm d}$ with the upper limit obtained by VISTA~\citep{Ackley:2020qkz}. The upper limit to the {\it K}-band emission provides approximately the same or weaker constraint on $M_{\rm d}$ than that to the {\it z}-band for $\theta_{\rm obs}\lesssim45^\circ$, but slightly tighter constraint for $\theta_{\rm obs}\gtrsim45^\circ$ due to weaker viewing angle dependence. For the model taking the fission fragments into account, $M_{\rm d}\geq0.03 M_\odot$ and 0.04 $M_\odot$ are disfavored for $\theta_{\rm obs}\lesssim50^\circ$ and $70^\circ$, respectively. This indicates that the observation in the near infrared wavelength is useful to detect or constraint the kilonova emission for a large viewing angle.

\begin{figure*}
 	 \includegraphics[width=.95\linewidth]{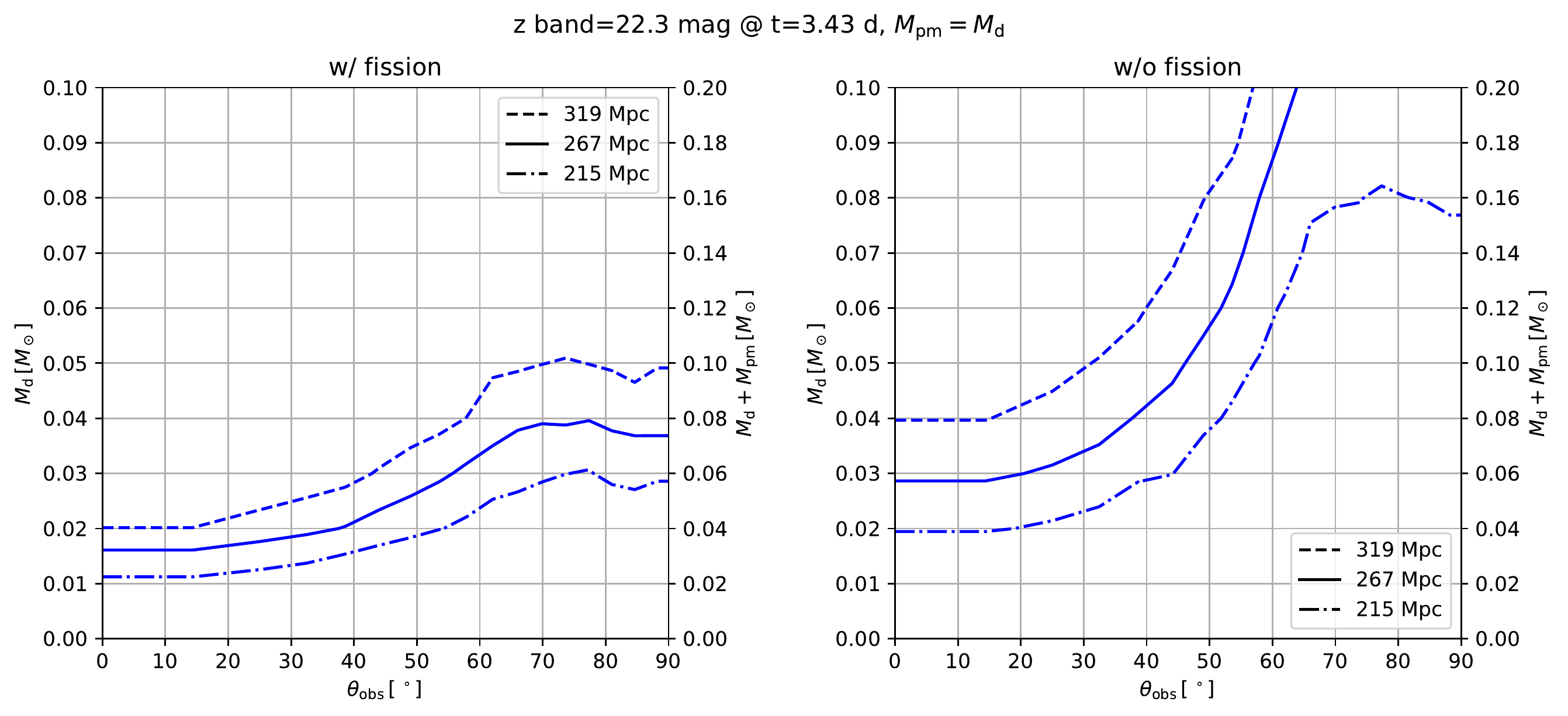}\\
 	 \includegraphics[width=.95\linewidth]{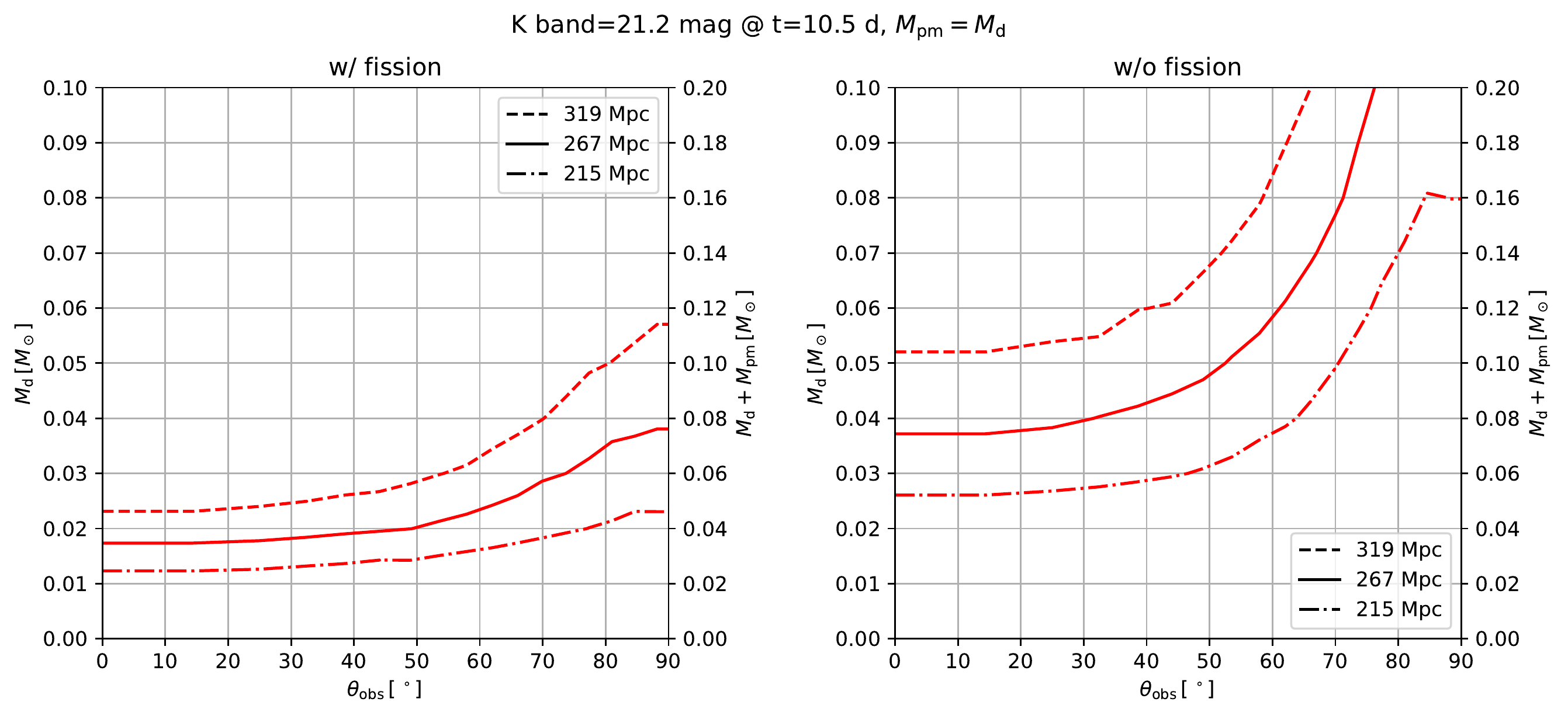}
	 \caption{Upper limit to the dynamical ejecta mass, $M_{\rm d}$, as a function of viewing angle, $\theta_{\rm obs}$, consistent with the upper limits. The top and bottom panels denote the results for the upper limits to the {\it z}-band emission at 3.43 d obtained by DECam~\citep{Andreoni:2019qgh} and the {\it K}-band emission at $t=10.5\,{\rm d}$ obtained by VISTA~\citep{Ackley:2020qkz}, respectively. The left and right panels show the results that take into account and omit the contribution from the fission fragments to the heating rate, respectively.  The dashed, solid, and dash-dotted curves denote the upper limits to the dynamical ejecta mass obtained assuming $D=319$ Mpc, 267 Mpc, and 215 Mpc, respectively.}
	 \label{fig:mag_th-meje}
\end{figure*}

As a summary, Figures~\ref{fig:mag_th-meje} shows the upper limit to the dynamical ejecta mass as a function of viewing angle, $\theta_{\rm obs}$. The figure shows that the upper limit to the dynamical ejecta mass is weaker by a factor of $\approx 2$--$3$ by omitting the contribution of the fission fragments to the heating rate. Overall, the constrains are not strong enough to indicate particular parameters of the binary.

\section{Discussion}\label{sec:discussion}
\subsection{Comparison with the ejecta mass constraints obtained in \cite{Andreoni:2019qgh} and~\cite{Ackley:2020qkz}} 

In the work of \cite{Andreoni:2019qgh}, the upper limit to the ejecta mass is obtained by employing the 1d kilonova models of~\cite{Hotokezaka:2019uwo} and 2d kilonova models of~\cite{Bulla:2019muo,Dhawan:2019phb}.~\cite{Ackley:2020qkz} also provides the upper limit to the ejecta mass based on the 2d analytical model of~\cite{Barbieri:2019sjc}. In this subsection we compare their results with ours.

First, we compare our results with those obtained by the 2d kilonova model of~\cite{Bulla:2019muo,Dhawan:2019phb}. For their model, (i) the ejecta density profile is simplified by a homologously expanding spherical ejecta distributing up to $0.3\,c$. (ii) The spherical ejecta is divided into the polar and  equatorial regions by certain degrees of latitude, and opacity models mimicking the lanthanide-poor and lanthanide-rich material are arranged in these regions, respectively. (iii) The lightcurves are calculated using a wavelength-dependent Monte Carlo code while the power-law temperature evolution as well as the time evolution of the opacity are assumed a priori and uniformly over the whole ejecta. On the other hand, the lanthanide-rich dynamical and post-merger ejecta with the density profile consistent with the numerical-relativity simulations~\citep[e.g.,][]{Foucart:2014nda,Kyutoku:2015gda,Foucart:2015vpa,Kyutoku:2017voj,Foucart:2019bxj,Metzger:2014ila,Wu:2016pnw,Siegel:2017nub,Siegel:2017jug,Fernandez:2018kax,Christie:2019lim,Fujibayashi:2020qda} are employed, and the temperature and opacity are evolved consistently with the radiative transfer in our model.

 Interestingly, regardless of the difference in the setups and the radiative transfer codes, the upper limit to the total ejecta mass is similar to that in \cite{Andreoni:2019qgh} for those omitting the contribution from the fission fragments\footnote{We note that the analytical model of heating rate employed in~\cite{Bulla:2019muo} approximately agrees with the heating rate employed in our models for the case that the contribution from the fission fragments is omitted.}. Indeed, the upper limits obtained for $M_{\rm pm}=M_{\rm d}$ shown in the lower right panel of Figure~\ref{fig:mag_th-meje} agree approximately with the upper limits obtained in~\cite{Andreoni:2019qgh}. However, we note that the agreement of the results may be a coincidence due to the facts that the model of~\cite{Bulla:2019muo} predicts fainter emission than our simulation for the same setup of ejecta~\citep{mbprivate}, while the lanthanide-poor ejecta ($Y_e=0.3$--$0.4$) arranged in the polar region of the ejecta model in~\cite{Bulla:2019muo} enhances the {\it z}-band emission at 3.43 d by $\approx0.5\,{\rm mag}$ (see Appendix~\ref{apx:YHcomp}).

Second, we compare our results with those of~\cite{Andreoni:2019qgh} obtained by employing the 1d kilonova model of \cite{Hotokezaka:2019uwo}. In the model of~\cite{Hotokezaka:2019uwo}, (i) a homologously expanding spherical ejecta with a single power-law density profile is employed,  (ii) the opacity is given by a constant value under the gray approximation, and (iii) the lightcurves are calculated based on the variant of Arnett's analytical model~\citep{1982ApJ...253..785A} with the stratified structure of the ejecta. The radioactive heating as well as its thermalization is computed based on a nuclear database and by taking the dependence on the decay energy into account (see~\cite{Hotokezaka:2019uwo} for the detail). In~\cite{Andreoni:2019qgh}, the models with the ejecta profile of $\rho\propto v^{-4.5}$ distributing from $0.1\,c$ to $0.4\,c$ are employed varying the value of gray opacity applied for the entire ejecta. 

The upper limit to the ejecta mass in the 1d kilonova models of~\cite{Hotokezaka:2019uwo} is weaker than that we obtained. While our results constrain the ejecta mass to be smaller than 0.06--0.07 $M_\odot$ for $D=215$ Mpc (see Figure~\ref{fig:magpm}), the ejecta mass is totally not constrained up to $0.1\,M_\odot$ by the 1d model in~\cite{Andreoni:2019qgh} for a typical value of opacity for lanthanide-rich ejecta ($\kappa\sim10 {\rm cm^2/g}$;~\citealt{Tanaka:2013ana}). This may be due to the low effective temperature of the emission which results from the high velocity edge of the ejecta profile in their model. The high velocity edge of the ejecta is set to be $0.4\,c$ in their model while $0.1\,c$ is employed for our post-merger ejecta model. The photosphere is located at a larger radius for such a model with high velocity edge, and the optical emission is suppressed because the spectra are redden. Indeed, we performed a calculation for our post-merger ejecta model with the maximum velocity twice as large value as the fiducial setup, that is, $0.2\,c$  (see Equation~(\ref{eq:dens})). We found that the {\it z}-band emission fainter by more than $1\,{\rm mag}$ is realized at 3.43 d for this model. Thus, while we employ conservative setups based on theoretical predictions obtained by numerical-relativity simulations, we could note that the constraint on the ejecta mass should depend largely on the assumptions on ejecta profiles and microphysical models employed.

Finally, we compare our results with those of~\cite{Ackley:2020qkz} obtained by employing the 2d kilonova model of \cite{Barbieri:2019sjc}.~\cite{Barbieri:2019sjc} consider the model with multiple ejecta components composed of the dynamical ejecta with non-spherical geometry and the post-merger ejecta with a spherical and equatorial-dominated density profile ($\propto {\rm sin}^2\theta$). The opacity of each ejecta component is given by a constant value under the gray approximation, while $15\,{\rm cm^2/g}$ and $5\,{\rm cm^2/g}$ are employed for the dynamical and post-merger components, respectively. Then the luminosity is calculated by determining the ejecta region from which photons can diffuse out, which is the extension of the methods introduced by~\cite{Piran:2012wd} and \cite{Kawaguchi:2016ana}.  The dynamical and post-merger ejecta are discretized in the radial and latitudinal cells, respectively, so that the viewing angle dependence of the lightcurves can be taken into account. The spectra are calculated by integrating the photon contribution from each discretized cell of ejecta in which blackbody emission is assumed. $\theta_{\rm obs}=30^\circ$ is assumed in their analysis.

~\cite{Ackley:2020qkz} conclude that $M_{\rm d}\geq0.1\,M_\odot$ is excluded with high confidence, which is consistent with our results of $\theta_{\rm obs}\leq45^\circ$. They also show that $M_{\rm d}\geq0.01\,M_\odot$ and $M_{\rm pm}\geq0.1\,M_\odot$ are disfavored at approximately 1$\sigma$ confidence. Although their results are broadly consistent with ours, they give a slightly tighter constraints to the ejecta mass. We suspect that a relatively small value of opacity ($5\,{\rm cm^2/g}$) employed for the post-merger ejecta is responsible for this difference (see Appendix~\ref{apx:YHcomp}). This indicates that employing a realistic setup for the ejecta opacity based on the  element abundances of numerical relativity simulations and atomic line opacity data taking its wavelength and density/temperature dependence is important to suppress the bias in the ejecta parameter estimation. 

\subsection{Implication to the future observation} 
\begin{figure*}
 	 \includegraphics[width=.95\linewidth]{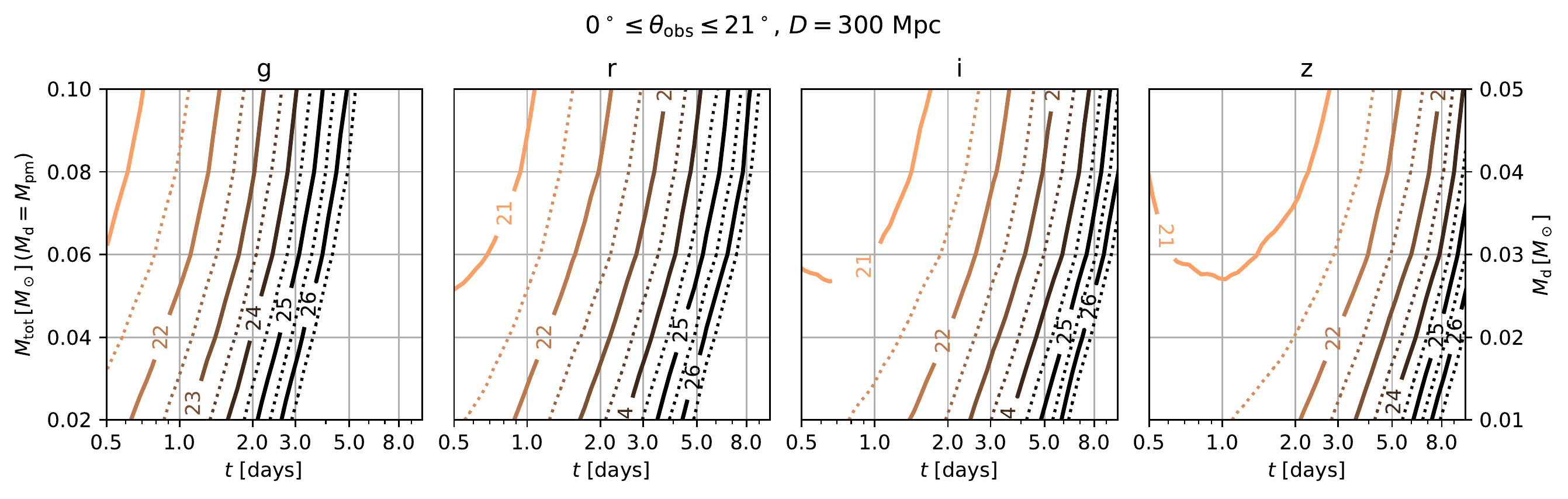}\\
 	 \includegraphics[width=.95\linewidth]{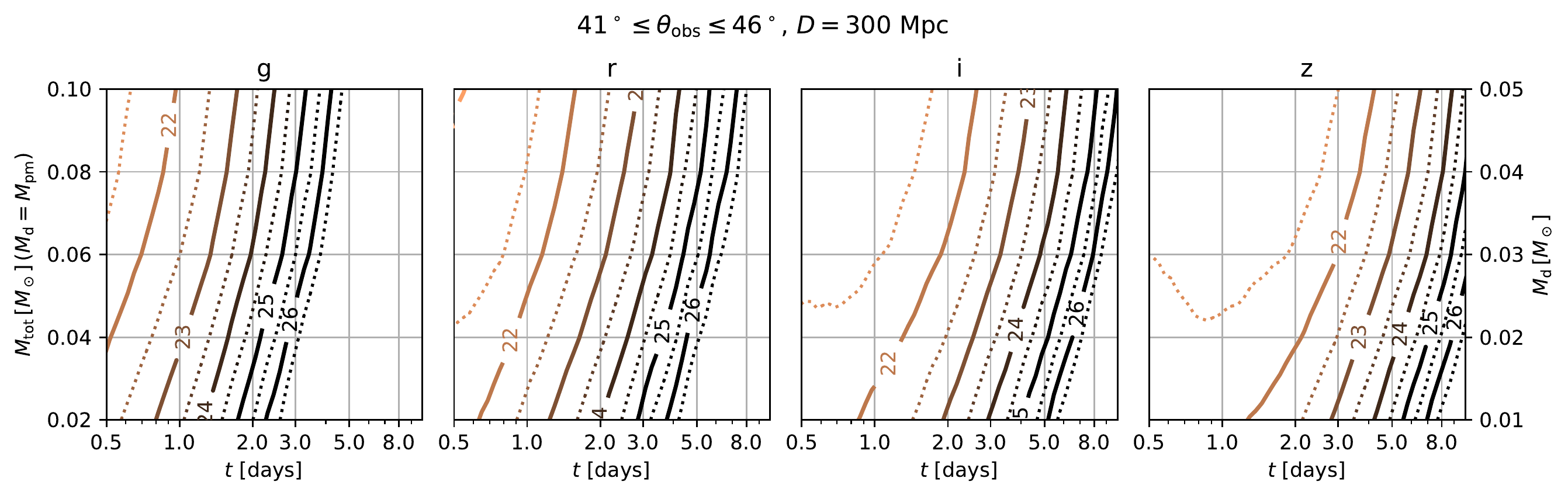}\\
 	 \includegraphics[width=.95\linewidth]{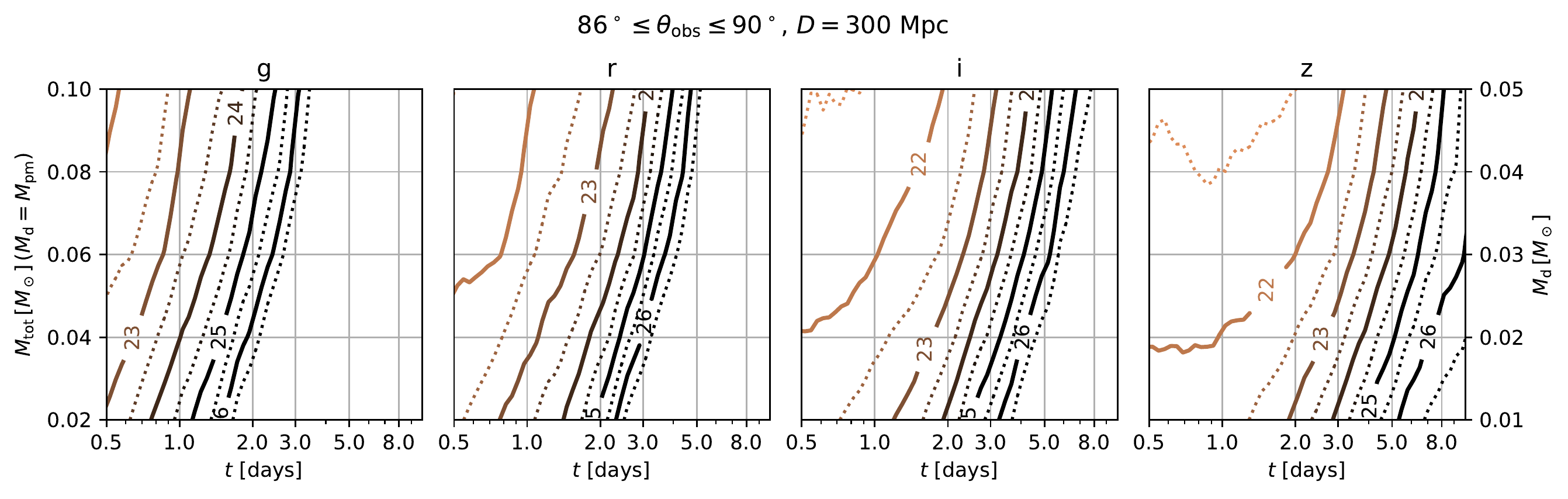}
	 \caption{Required depth of the observation in the {\it griz}-band filters to detect the BH-NS kilonovae with $M_{\rm pm}=M_{\rm d}$. The top, middle, and bottom panels denote the results for $0^\circ\le\theta_{\rm obs}\le20^\circ$, $41^\circ\le\theta_{\rm obs}\le46^\circ$, and $86^\circ\le\theta_{\rm obs}\le90^\circ$, respectively. The hypothetical distance to the event is set to be $300$ Mpc. The dotted curves denote the contours with 0.5 mag interval. We note that the results before $t=1\,{\rm d}$ may be not very reliable due to lack of the opacity table for highly ionized atoms (see \cite{Tanaka:2019iqp}).}
	 \label{fig:mag_t-meje}
\end{figure*}
\begin{figure*}
 	 \includegraphics[width=.95\linewidth]{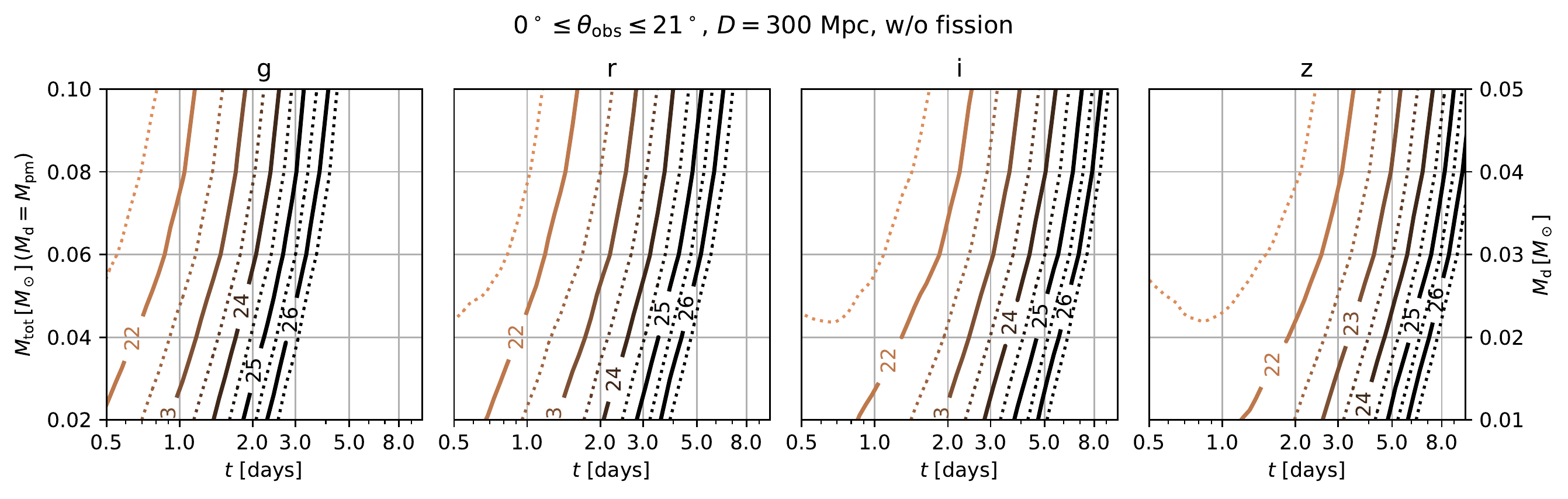}\\
 	 \includegraphics[width=.95\linewidth]{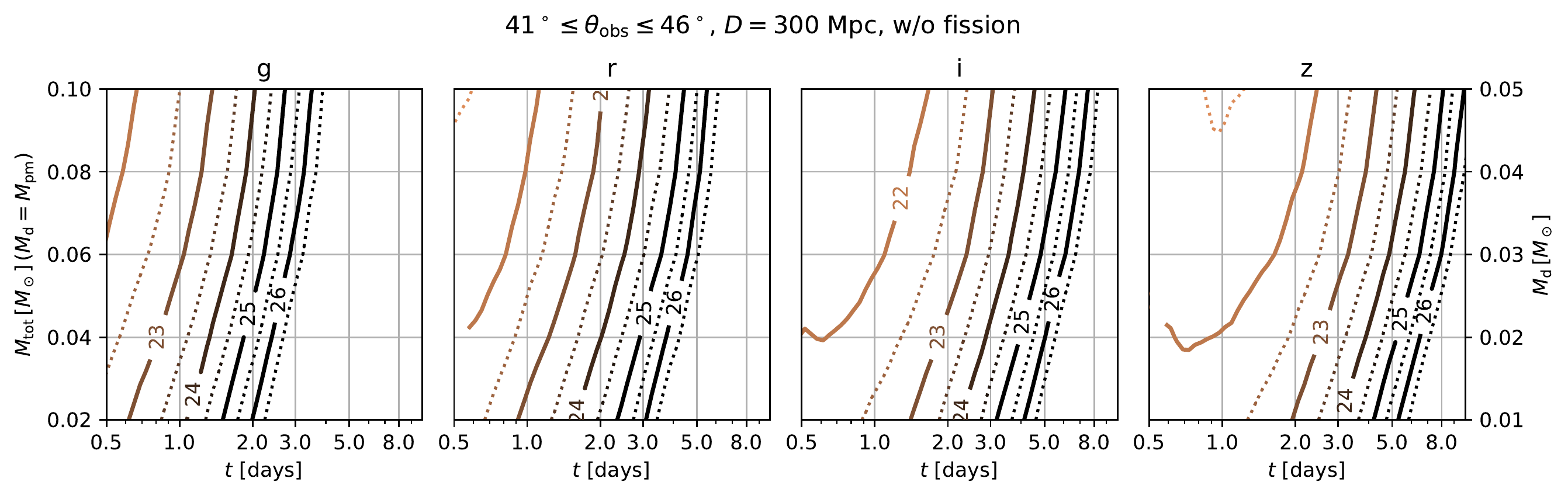}\\
 	 \includegraphics[width=.95\linewidth]{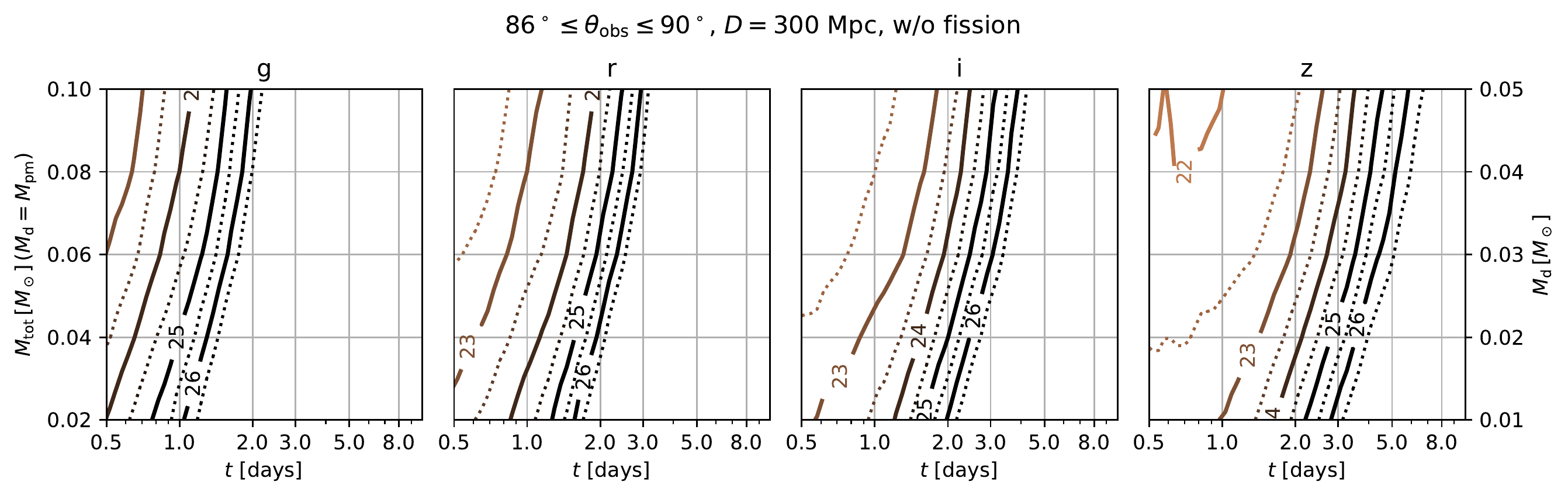}
	 \caption{The same as for Figure~\ref{fig:mag_t-meje} but for the models without fission fragments.}
	 \label{fig:mag_t-meje_nofis}
\end{figure*}
Figures~\ref{fig:mag_t-meje} and~\ref{fig:mag_t-meje_nofis} show the required depth of the observation in the {\it griz}-band filters to detect the BH-NS kilonovae for $M_{\rm pm}=M_{\rm d}$ at $D=300$\,Mpc with and without the contribution of fission fragments to the heating rate, respectively. Irrespective of the fission fragments, the emission becomes brighter at longer wavelengths, and hence, a kilonova with less ejecta mass can be  observed if the observations are performed in a longer wavelength (e.g., the {\it z}-band rather than the {\it g}-band). Also, a kilonova with the same ejecta mass can be observed in the later epoch by the same depth of the observation in a band filter with longer wavelength. Thus, the observation in the ${\it i}$ or ${\it z}$ band could be useful to detect the kilonova~\citep{Barnes:2013wka,Tanaka:2013ixa}.

Focusing on the case that the event is observed from the polar direction ($\theta_{\rm obs}\leq 45^\circ$), the follow-up observation deeper than $22$ mag within $2$ d is crucial to detect the kilonovae of $M_{\rm pm}=M_{\rm d}=0.03\,M_\odot$. On the other hand, if the event is observed from the equatorial direction ($\theta_{\rm obs}\geq 70^\circ$), the observation deeper than $23$ mag within $2$ d is required.

\subsection{Constrain on the NS mass-radius relation} 
\begin{figure*}
 	 \includegraphics[width=.48\linewidth]{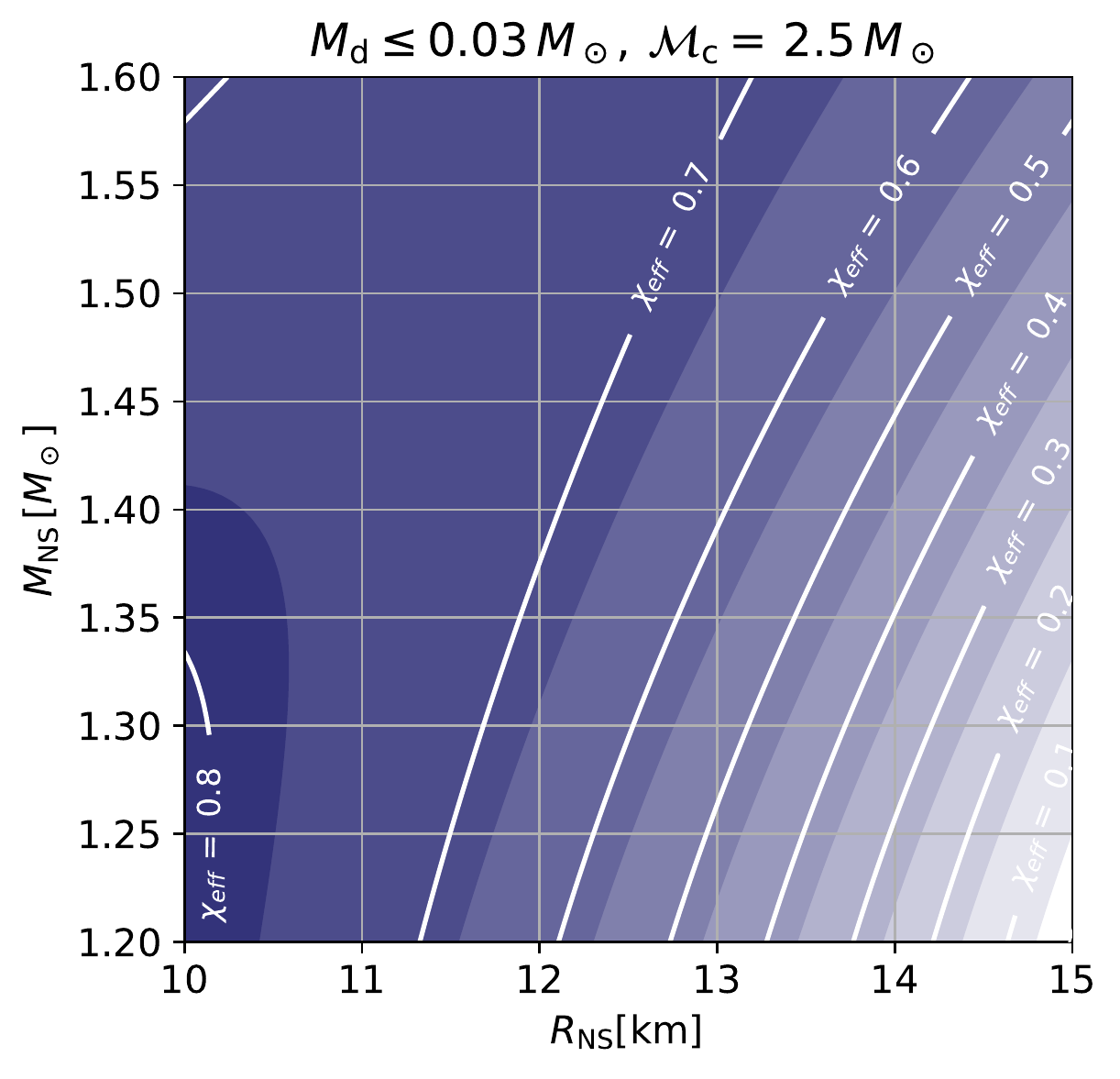}
 	 \includegraphics[width=.48\linewidth]{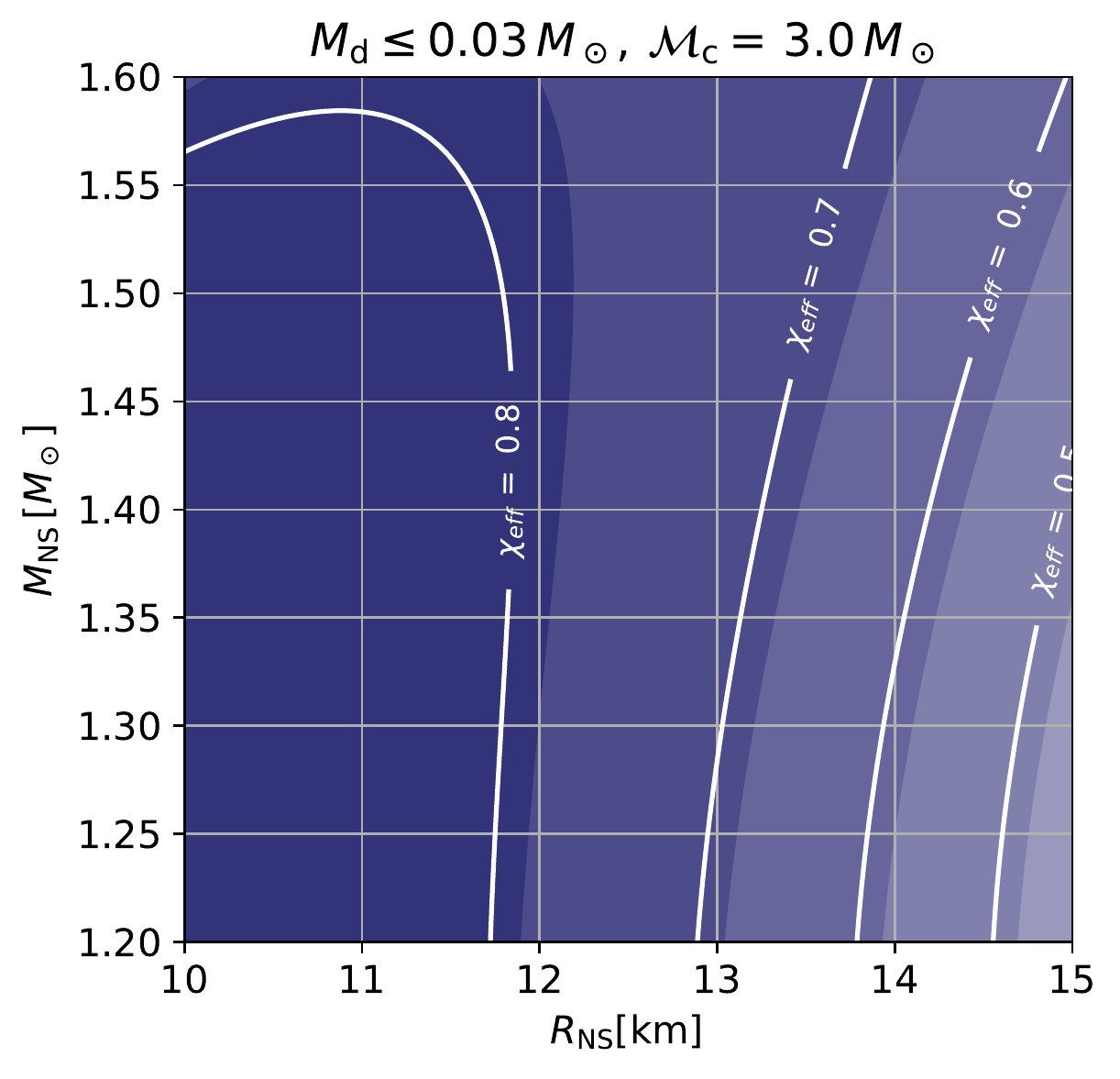}
	 \caption{The allowed region of the NS mass and radius for $M_{\rm d}\le0.03\,M_\odot$ with ${\cal M}_{\rm c}=2.5\,M_\odot$ and $3.0\,M_\odot$. Each white curve with the effective spin value denotes the NS mass and radius for which $M_{\rm d}=0.03\,M_\odot$ is predicted by the fitting formula~\citep{Kawaguchi:2016ana,Coughlin:2017ydf} using the corresponding value of the effective spin. The NS mass and radius are allowed only in the left side of the curve for a given upper limit to the effective spin. We note that the boundaries of the deeper-color regions are determined by $M_{\rm d}-\Delta M_{\rm d}=0.03\,M_\odot$ to take the estimated error of the fitting formula, $\Delta M_{\rm d}$, into account (see~\cite{Kawaguchi:2016ana} for the detail).}
	 \label{fig:const_rns_eje003}
\end{figure*}

\begin{figure*}
 	 \includegraphics[width=.48\linewidth]{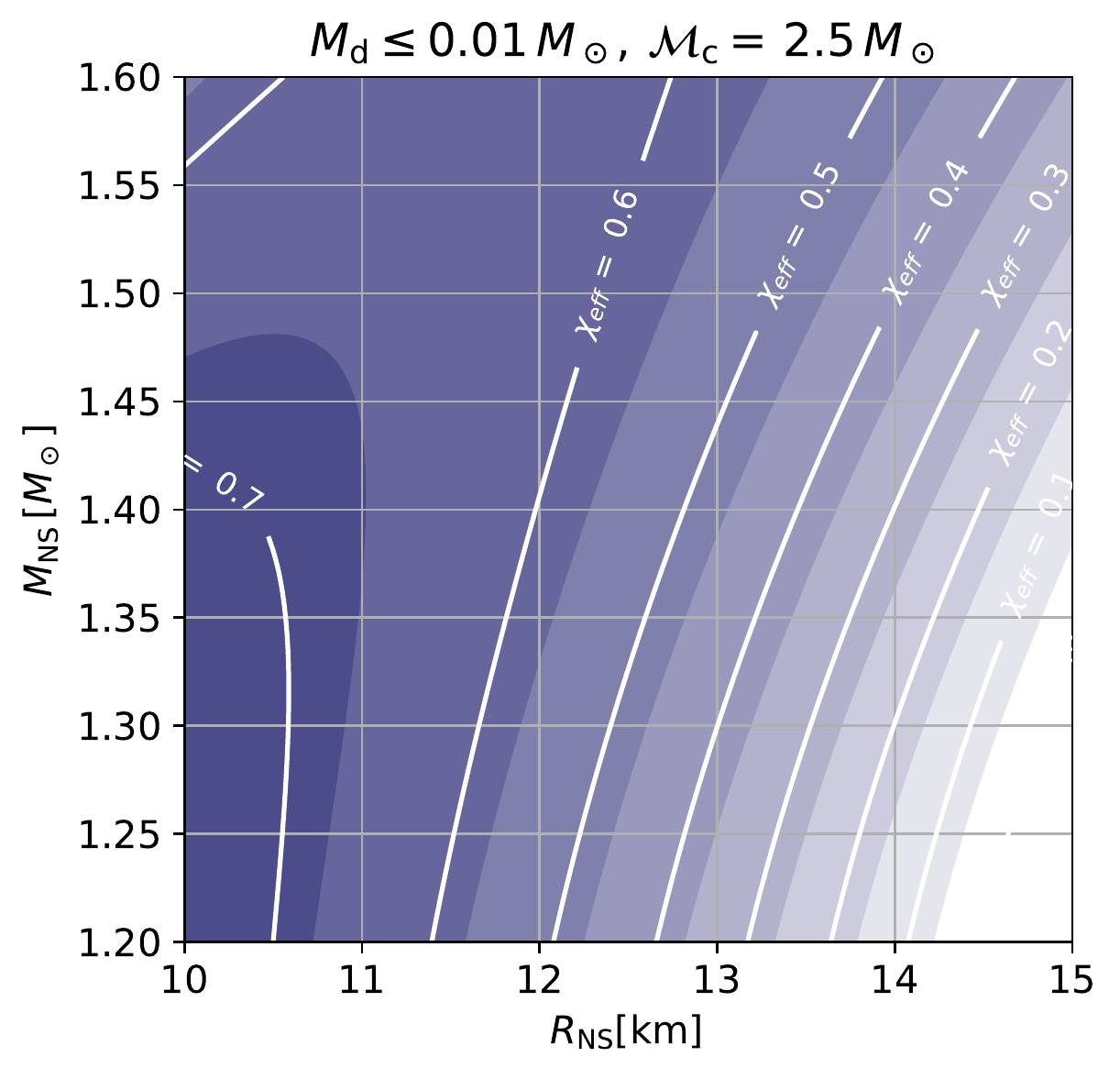}
 	 \includegraphics[width=.48\linewidth]{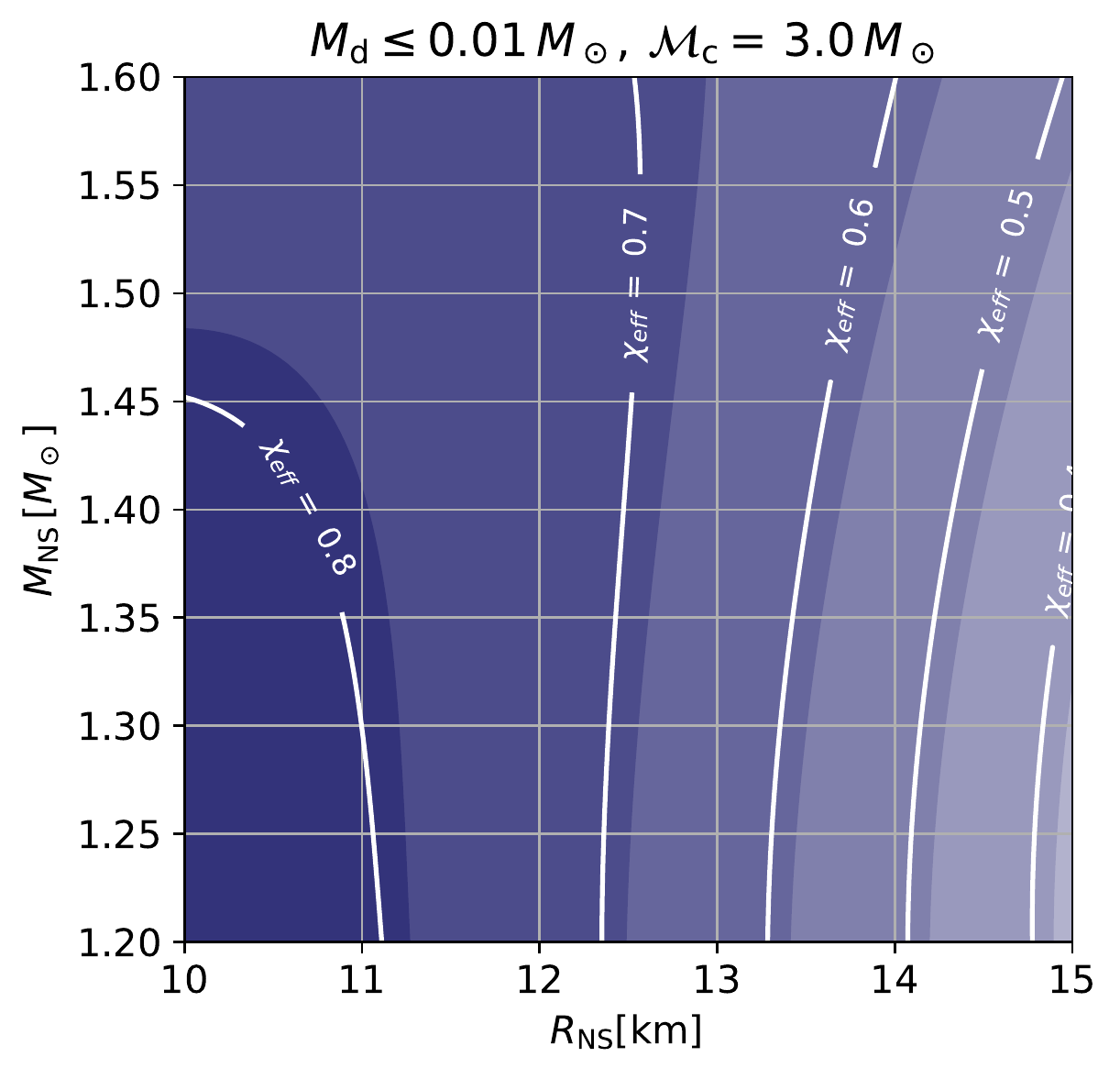}
	 \caption{The same as Figure~\ref{fig:const_rns_eje003} but for $M_{\rm d}\leq0.01\,M_\odot$.}
	 \label{fig:const_rns_eje001}
\end{figure*}

By combining binary parameters inferred by the GW data analysis, the constraint on the dynamical ejecta mass can be used to constrain the mass-radius relation of a NS. The chirp mass of the binary, defined by ${\cal M}_{\rm c}=\frac{M_{\rm BH}^{3/5}M_{\rm NS}^{3/5}}{\left(M_{\rm BH}+M_{\rm NS}\right)^{1/5}}$ with the BH mass $M_{\rm BH}$ and the NS mass $M_{\rm NS}$, is a quantity which can be determined most accurately from the GW data analysis. Also the mass ratio, $q=M_{\rm BH}/M_{\rm NS}$, as well as the so-called effective spin, $\chi_{\rm eff}=\frac{M_{\rm BH}}{M_{\rm BH}+M
_{\rm NS}}\chi_{\rm BH}$, are measured for some extent. Here, $\chi_{\rm BH}$ denotes the component of the dimensionless BH spin parallel to the orbital axis, and we assume that the NS spin is negligible~\citep{Burgay:2003jj,Tauris:2017omb,Abbott2017}. The previous numerical simulations for BH-NS mergers show that the dynamical ejecta mass is determined approximately by these parameters and the NS radius~\citep{Kawaguchi:2016ana}. Thus, the constraint on the dynamical ejecta mass could be translated to the constraint on the NS radius if the parameters introduced above are determined by the GW data analysis. We note that the similar analysis was already performed by~\cite{Coughlin:2019zqi} and~\cite{Andreoni:2019qgh} focusing on the total ejecta mass, but our analysis focuses on constraining the NS mass-radius relation based on the parameters which can be obtained directly by the GW data analysis.

Employing the analytical fit of the dynamical ejecta mass~\citep{Kawaguchi:2016ana,Coughlin:2017ydf}, we calculate the allowed region of the NS mass and radius for a given chirp mass, an upper limit to the dynamical ejecta mass, and a lower limit to the effective spin. We plot the results for $M_{\rm d}\le0.03\,M_\odot$ in Figure~\ref{fig:const_rns_eje003} as an illustration. The left and right panels in Figure~\ref{fig:const_rns_eje003} show the cases for ${\cal M}_{\rm c}=2.5\,M_\odot$ and $3.0\,M_\odot$, respectively. The NS mass and radius are constrained to the region in the contour for a given lower limit of the effective spin for $M_{\rm d}\le0.03\,M_\odot$. Primarily, this analysis provides the upper limit to the NS radius for given NS mass because $M_{\rm d}$ becomes large as the NS radius increases. We note that the fitting formula for the dynamical ejecta mass employed here is calibrated to the numerical-relativity simulations only for $M_{\rm NS}\approx1.4\,M_\odot$ and $4 M_\odot\lesssim M_{\rm BH} \lesssim 10 M_\odot$, i.e., only for ${\cal M}_{\rm c} \approx 2.0$--$3.0 M_\odot$~\citep{Kawaguchi:2016ana}.

A tighter constraint is given for a larger upper limit to the effective spin because the dynamical ejecta mass increases for the case that the BH is more rapidly spinning. The constraint on the NS mass and radius becomes weaker as the chirp mass of the binary increases. This reflects the fact that the BH mass is a monotonically increasing function of the chirp mass for a fixed NS mass and the dynamical ejecta mass decreases as the BH mass increases in the range of ${\cal M}_{\rm c}$ and $M_{\rm NS}$ shown in Figure~\ref{fig:const_rns_eje003}.\footnote{We note that the dynamical ejecta mass would be an increasing function of the BH mass for a small mass ratio $q\lesssim 3$~\citep{Foucart:2019bxj}, while $q$ is always larger than 3 in the range of ${\cal M}_{\rm c}$ and $M_{\rm NS}$ shown in Figure~\ref{fig:const_rns_eje003}.} 

For ${\cal M}_{\rm c}=2.5\,M_\odot$ (e.g., for $M_{\rm NS}=1.4\,M_\odot$ and $M_{\rm BH}=6.5\,M_\odot$), the condition of $\chi_{\rm eff} \geq 0.6$ gives a meaningful upper limit to the NS radius ($<12$--$14\,{\rm km}$). On the other hand, for ${\cal M}_{\rm c}=3\,M_\odot$  (e.g., for $M_{\rm NS}=1.4\,M_\odot$ and $M_{\rm BH}=10\,M_\odot$), the NS radius can be constrained at most $\lesssim14\,{\rm km}$ even if the effective spin is inferred to be larger than $0.6$. Thus, a BH-NS event with the chirp mass smaller than $\lesssim 3\,M_\odot$ and effective spin larger than $\gtrsim 0.5$ would be important for providing the constraint on the NS mass-radius relation.

A tighter constraint on the NS mass-radius relation can be obtained if the dynamical ejecta mass is constrained to be a smaller value. Figure~\ref{fig:const_rns_eje001} shows the same as Figure~\ref{fig:const_rns_eje003} but for the case that the dynamical ejecta is constrained to be less than $0.01\,M_\odot$. For example, the NS radius typically smaller by $\approx0.5\,{\rm km}$ is allowed than that for $M_{\rm d}\leq0.03\,M_\odot$ for a given value of lower limit to $\chi_{\rm eff}$. As is the case for $M_{\rm d}\leq0.03\,M_\odot$, the NS radius up to $14\,{\rm km}$ is always allowed for ${\cal M}_{\rm c}=3\,M_\odot$ unless $\chi_{\rm eff}$ is inffered to be larger than $0.5$. Thus, a BH-NS event with ${\cal M}_{\rm c}\lesssim 3\,M_\odot$ is also crucial to obtain a valuable constrain to the NS mass-radius relation with the upper limit of $M_{\rm d}\leq0.01\,M_\odot$.

More stringent constraint on the NS mass and radius can be obtained for the case that the kilonova of a BH-NS event is observed and the range of the dynamical ejecta mass is restricted. With both upper and lower limits to the dynamical ejecta mass and effective spin, the upper and lower limits to NS radius can be obtained presuming the accurate measurement of the chirp mass. For example, if the dynamical ejecta mass is suggested to be larger than $0.01\,M_\odot$, in addition to the constraint obtained by the upper limit to the dynamical ejecta mass, the NS mass and radius are restricted in the region in Figure~\ref{fig:const_rns_eje001} where the effective spin is smaller than the upper limit inferred by the GW analysis. Figures~\ref{fig:const_rns_eje003} and~\ref{fig:const_rns_eje001} indicate that, to constrain the NS radius within $\approx1\,{\rm km}$ error, constraints on the dynamical ejecta mass and effective spin with $\Delta M_{\rm d}\lesssim0.01\,M_\odot$ and $\Delta \chi_{\rm eff} \lesssim0.1$ for the BH-NS event of ${\cal M}_c\leq3.0\,M_\odot$ are crucial.

\section{Summary}\label{sec:summary}
In this paper, we studied the upper limit to the ejecta mass based on the upper limits to the emission obtained by the EM counterpart followup campaigns for the BH-NS merger candidate event S190814bv by performing radiative transfer simulations for kilonovae. In our calculation, the realistic ejecta density profile as well as the detailed opacity and heating rate models consistent with the numerical-relativity simulations~\citep[e.g.,][]{Foucart:2014nda,Kyutoku:2015gda,Foucart:2015vpa,Kyutoku:2017voj,Foucart:2019bxj,Metzger:2014ila,Wu:2016pnw,Siegel:2017nub,Siegel:2017jug,Fernandez:2018kax,Christie:2019lim,Fujibayashi:2020qda} are employed. In addition, the temperature and opacity are evolved consistently with the radiative transfer. In this study, we found that the upper limit to the {\it z}-band emission at $t=3.43$ d obtained by DECam~\citep{Andreoni:2019qgh} and the upper limit to the {\it K}-band at 9.2--10.5 d by VISTA~\citep{Ackley:2020qkz} give the tightest constraint on the kilonova lightcurve model.

We showed that the brightness of the emission in the {\it z}-band at $t=3.43$ d depends not only on the total ejecta mass but also on the ratio between the dynamical and post-merger ejecta mass. We showed that the model only with the post-merger ejecta gives the faintest emission for $\theta_{\rm obs}\lesssim45^\circ$ for given total ejecta mass, while the faintest emission for $\theta_{\rm obs}\gtrsim60^\circ$ is realized for the case that $20$--$50\%$ of the ejecta is the dynamical component. We also found that the {\it K}-band emission at $t=10.5$ d has broadly the same dependence, while the viewing angle dependence is weaker than that for the {\it z}-band emission. We found that the total ejecta mass larger than $0.1\,M_\odot$ is consistent with the upper limits to the {\it z} and {\it K} bands for $D\geq267\,{\rm Mpc}$ or for $\theta_{\rm obs}\gtrsim60^\circ$. For $\theta_{\rm obs}\leq45^\circ$ and $D=215\,{\rm Mpc}$, the total ejecta mass is constrained to be less than $0.07\,M_\odot$. However, these upper limits are not strong enough to indicate particular parameters of the binary. Thus, although there always exists a trade-off between the depth and the area, we recommend deeper observations than that in the current strategy to detect or to obtain a tight constraint on the kilonovae at $D\gtrsim 200\,{\rm Mpc}$. 


We also studied the upper limit to the ejecta mass focusing on the dynamical component. For the case that the post-merger ejecta mass is larger than the dynamical ejecta mass and taking the contribution of the fission fragments to the heating rate into account, we found that the dynamical ejecta mass has to be smaller than $0.02\,M_\odot$, $0.03\,M_\odot$, and $0.05\,M_\odot$ for $\theta_{\rm obs}\le 20^\circ$, $\theta_{\rm obs}\le 50^\circ$, and for the entire viewing angle, respectively. We also showed that the upper limit to the dynamical ejecta mass is affected strongly by the uncertainty in the contribution of the fission fragments to the heating rate. If the contribution of the fission fragments to the heating rate is omitted, the models with the dynamical ejecta mass as large as $0.05\,M_\odot$ is consistent with the upper limits to the {\it z} and {\it K} bands for $\theta_{\rm obs}\ge 30^\circ$.

In Figures~\ref{fig:mag_t-meje} and~\ref{fig:mag_t-meje_nofis}, we summarize the depth of observation required to detect the kilonova for given total ejecta mass for the cases with $M_{\rm pm}=M_{\rm d}$. We showed that, for the case that a BH-NS merger event is detected by GWs from the polar direction ($\theta_{\rm obs}\leq 45^\circ$) at $D=300$ Mpc, the {\it iz}-band observation deeper than $22$ mag within 2 d after the GW trigger is crucial to detect the kilonova with the total ejecta mass of $0.06\,M_\odot$ (and the dynamical ejecta of $0.03\,M_\odot$). To achieve this, the EM follow-up by 4/8-m class telescopes are crucial~\citep{Nissanke:2012dj}. We note that the kilonova detection will be more feasible in the presence of the lanthanide-poor post-merger ejecta particularly in the shorter wavelengths (e.g., in the {\it g,z}-bands, see also Appendix~\ref{apx:YHcomp}) ~\citep[e.g.][]{Metzger:2014ila,Tanaka:2017lxb,Kawaguchi:2019nju}.

We showed that the constraint on the dynamical ejecta mass can be used to constrain the mass-radius relation of a NS by combining the binary parameter inferred by the GW data analysis, such as the chirp mass and effective spin. We showed that a BH-NS event with the chirp mass smaller than $\lesssim 3\,M_\odot$ and effective spin larger than $\gtrsim 0.5$ can provide interesting indication to the NS mass-radius relation by this analysis if the dynamical ejecta mass $\lesssim 0.03\,M_\odot$ is obtained.

\begin{acknowledgments}
We thank Mattia Bulla for a valuable discussion and the cross-comparison of the radiative transfer simulation codes. Numerical computation was performed on Cray XC40 at Yukawa Institute for Theoretical Physics, Kyoto University and Sakura cluster at Max Planck Institute for Gravitational Physics (Albert Einstein Institute). This work was supported by Grant-in-Aid for Scientific Research (JP16H02183, JP16H06342, JP17H01131, JP15K05077, JP17K05447, JP17H06361, JP15H02075, JP17H06363, 18H05859) of JSPS and by a post-K computer project (Priority issue No. 9) of Japanese MEXT.
\end{acknowledgments}
\
\bibliographystyle{apj}
\bibliography{ref}
\appendix

\section{The upper limit to the dynamical ejecta mass for $M_{\rm pm}=0.5 M_{\rm d}$}\label{apx:mpm05md}

\begin{figure*}
 	 \includegraphics[width=.5\linewidth]{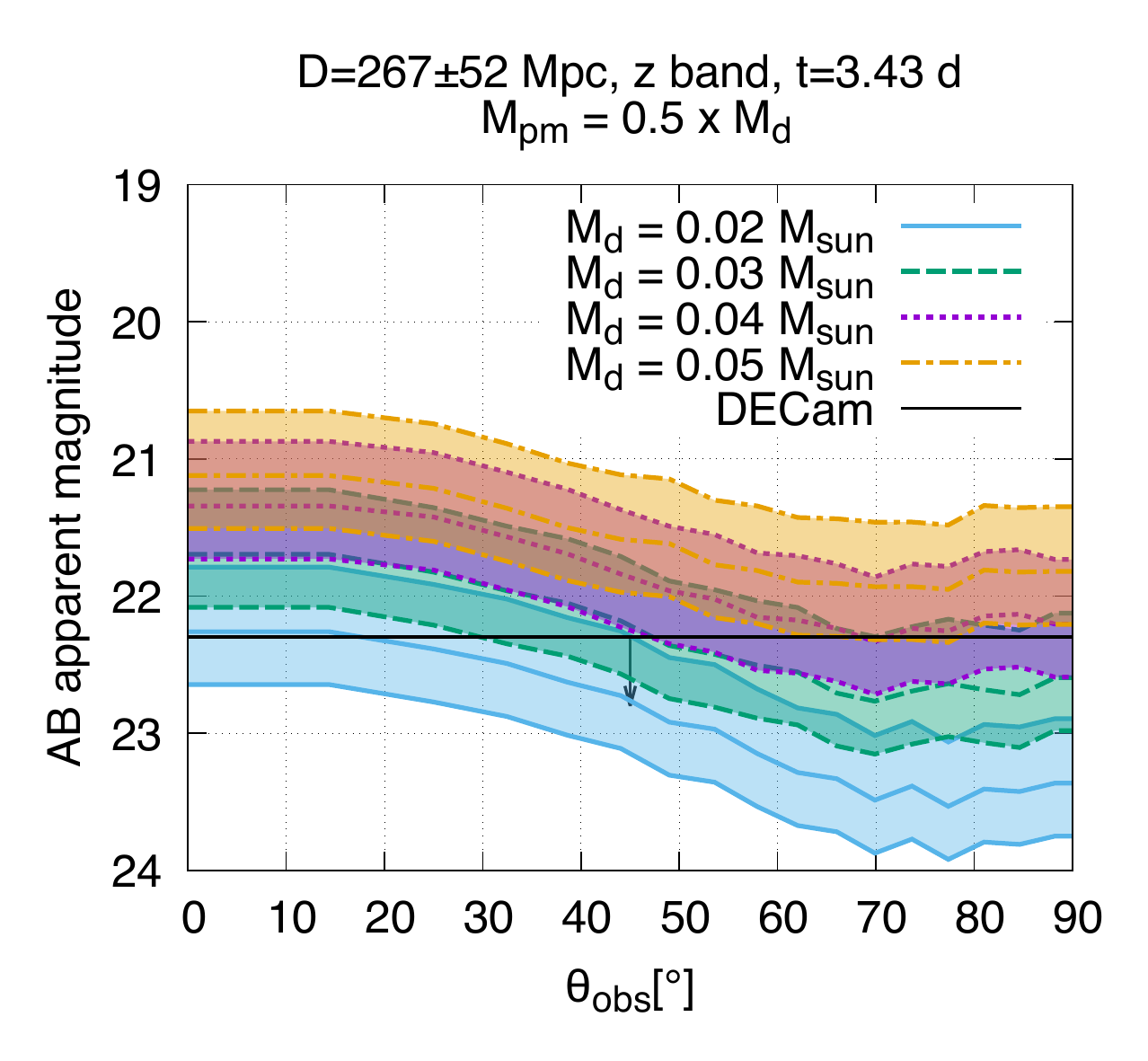}
 	 \includegraphics[width=.5\linewidth]{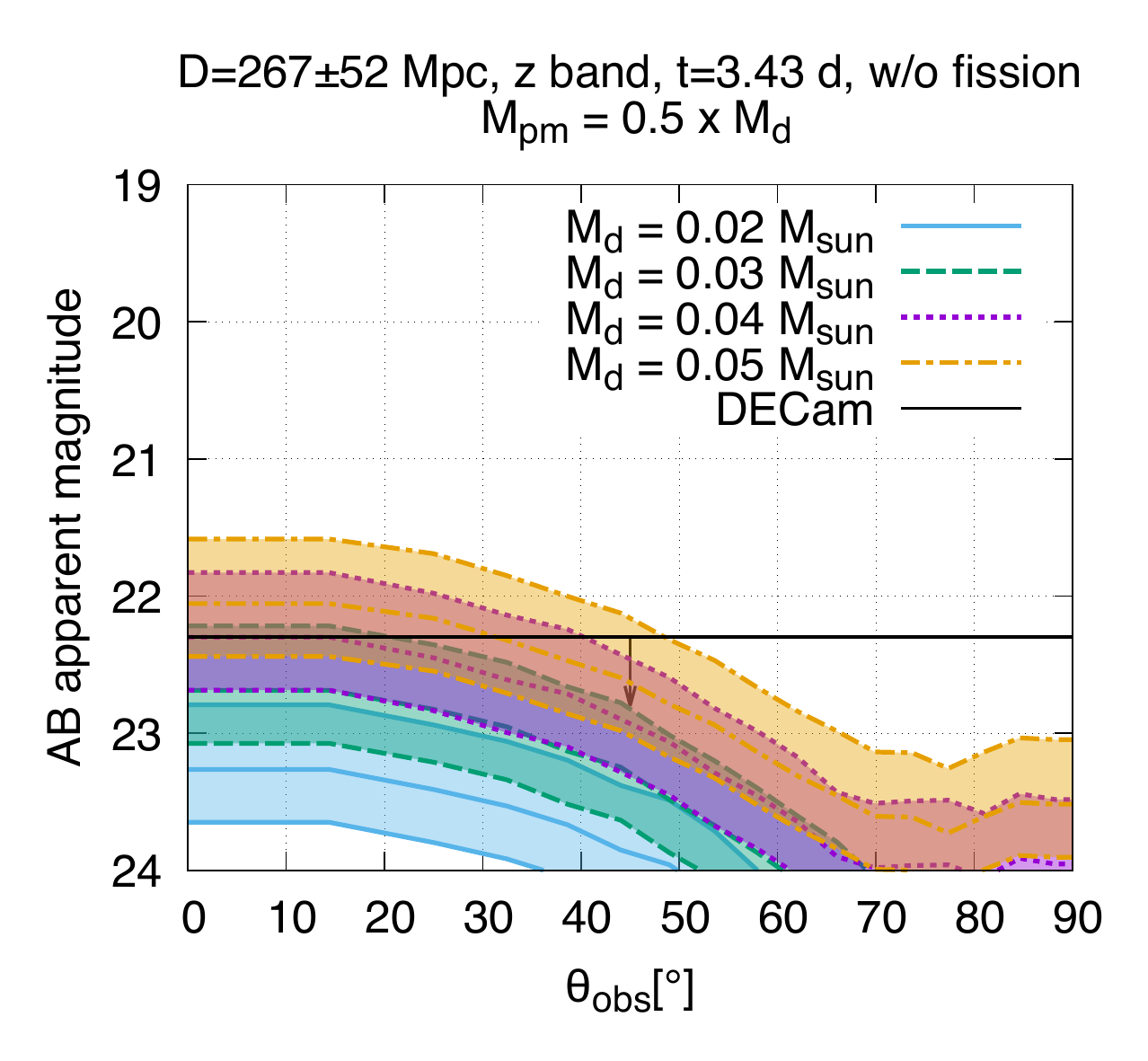}
	  	 \caption{The same as Figure~\ref{fig:mag_comp} but for the models with $(M_{\rm d},M_{\rm pm})=(0.02\,M_\odot, 0.01\,M_\odot)$ (blue solid), $(0.03\,M_\odot, 0.015\,M_\odot)$ (green dashed), $(0.04\,M_\odot, 0.02\,M_\odot)$ (purple dotted), and $(0.05\,M_\odot, 0.025\,M_\odot)$ (orange dotted). The right and left panels show the models in which the contribution from the fission fragments to the heating rate is taken into account and omitted, respectively.}
	 \label{fig:mag_comp-2}
\end{figure*}

\begin{figure*}
 	 \includegraphics[width=.5\linewidth]{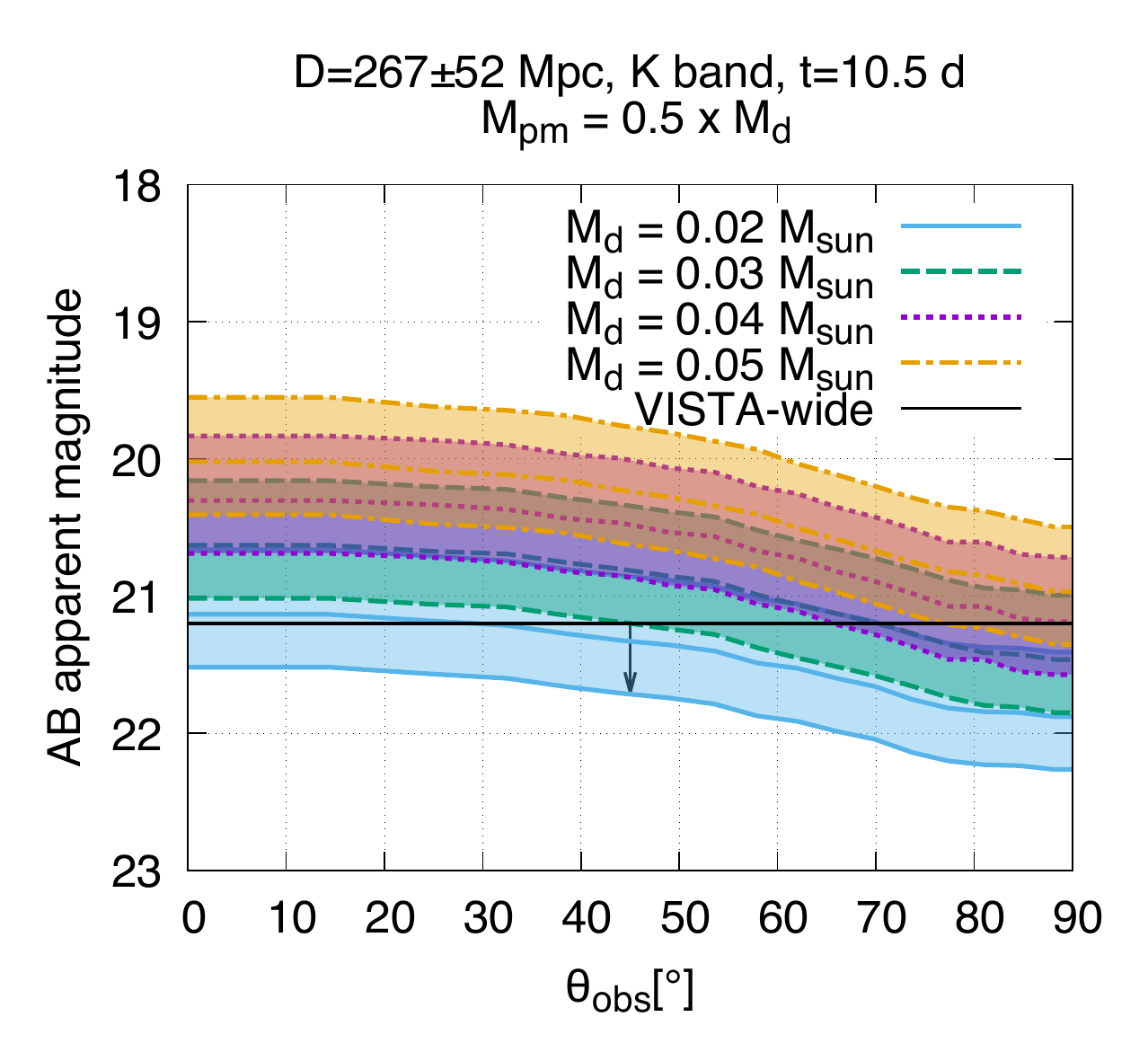}
 	 \includegraphics[width=.5\linewidth]{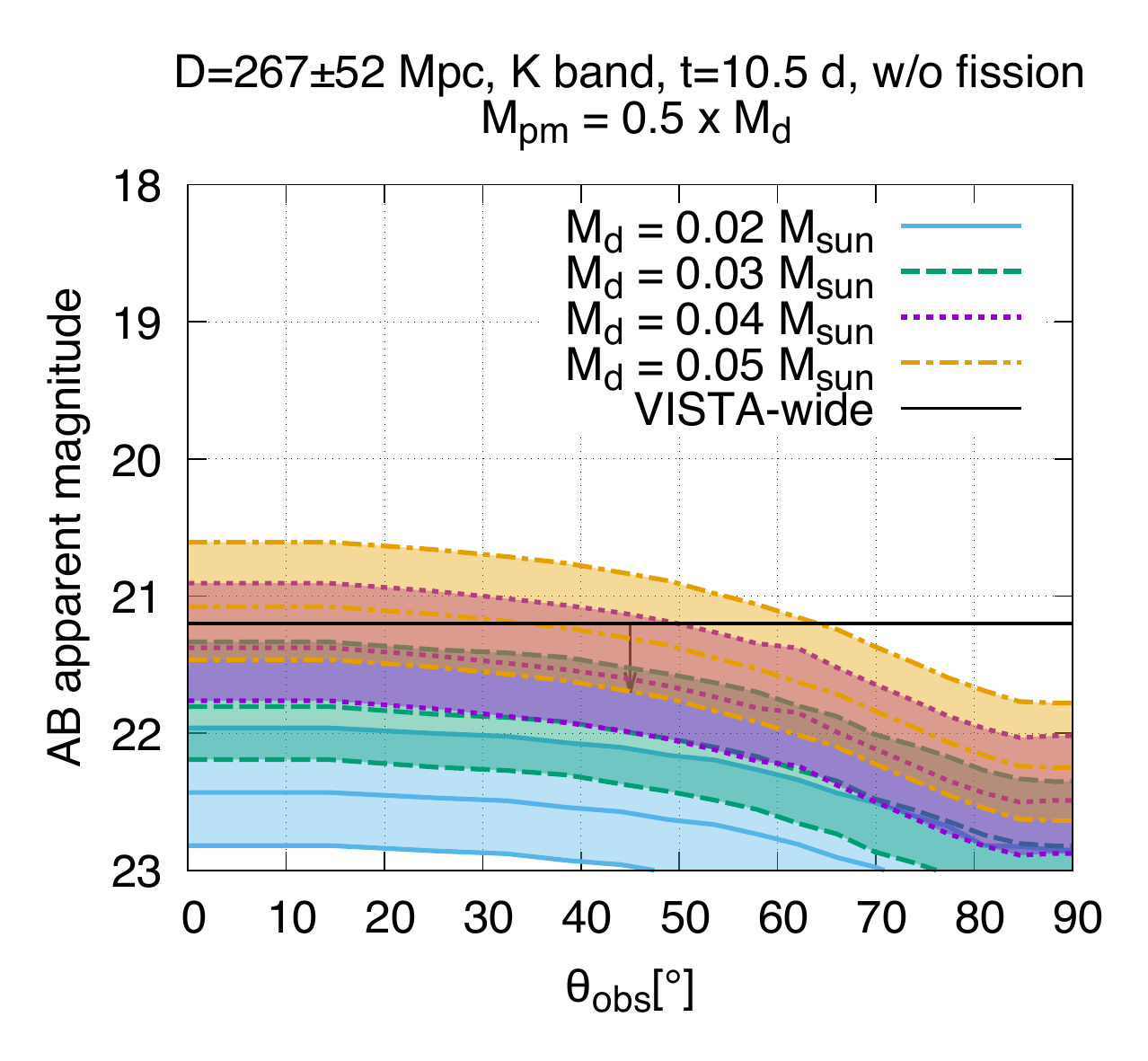}
	  	 \caption{The same as Figure~\ref{fig:mag_comp-2} but for the brightness of the {\it K}-band emission at $t=10.5\,{\rm d}$ with the upper limit obtained by VISTA~\citep{Ackley:2020qkz}.}
	 \label{fig:mag_comp-2_K}
\end{figure*}

The upper limit to the dynamical ejecta mass becomes weaker if we allow the post-merger ejecta mass to be smaller than the dynamical ejecta mass. Indeed, for some cases of BH-NS mergers, the remnant torus mass could be comparable to the dynamical ejecta mass (see the models labeled with Q7a5 in \cite{Kyutoku:2015gda}), and hence, the post-merger ejecta mass could be smaller than the dynamical one. Thus, we also explore the cases that the post-merger ejecta mass is the half of the dynamical ejecta mass ($M_{\rm pm}=0.5 M_{\rm d}$). 

Figure~\ref{fig:mag_comp-2} shows the same as Figure~\ref{fig:mag_comp} but for the models with $(M_{\rm d},M_{\rm pm})=(0.02\,M_\odot, 0.01\,M_\odot)$, $(0.03\,M_\odot, 0.015\,M_\odot)$, $(0.04\,M_\odot, 0.02\,M_\odot)$, and $(0.05\,M_\odot, 0.025\,M_\odot)$. First, we focus on the models in which the contribution from the fission fragments to the heating rate is taken into account (see the left panel). The left panel of Figure~\ref{fig:mag_comp-2} shows that the models with $2M_{\rm pm}=M_{\rm d} \geq0.03\,M_\odot$ and $0.04\,M_\odot$ are disfavored for $\theta_{\rm obs}\le 30^\circ$ and $\theta_{\rm obs}\le 50^\circ$, respectively. The models with $2M_{\rm pm}=M_{\rm d} >0.05\,M_\odot$ is disfavored for the entire viewing angle as is the case for the model with $M_{\rm pm}=M_{\rm d}$. We note that the brightness of the emission for $\theta_{\rm obs}\gtrsim 70^\circ$ is approximately the same as for the model with $M_{\rm pm}=M_{\rm d}$. This indicates that the emission for $\theta_{\rm obs}\gtrsim 70^\circ$ is dominated by the emission from the dynamical ejecta. 

The upper limit to the ejecta mass is weaker for the models without the fission fragments. The right panel of Figure~\ref{fig:mag_comp-2} shows that, even for the case of $2M_{\rm pm}=M_{\rm d}=0.05M_\odot$, the model is marginally consistent with the upper limit to the {\it z}-band emission at $t=3.43$ d for the entire viewing angle by omitting the contribution from the fission fragments to the heating rate.

Figure~\ref{fig:mag_comp-2_K} shows the same as Figure~\ref{fig:mag_comp-2} but for the brightness of the {\it K}-band emission at $t=10.5\,{\rm d}$ with the upper limit obtained by VISTA~\citep{Ackley:2020qkz}. As is the same as in Section~\ref{sec:results_dyn}, slightly weaker and tighter constraints on $M_{\rm d}$ are obtained for $\theta_{\rm obs}\lesssim45^\circ$ and for $\theta_{\rm obs}\gtrsim45^\circ$, respectively, by the upper limit to the {\it K}-band emission than that to the {\it z}-band.

\begin{figure*}
 	 \includegraphics[width=.95\linewidth]{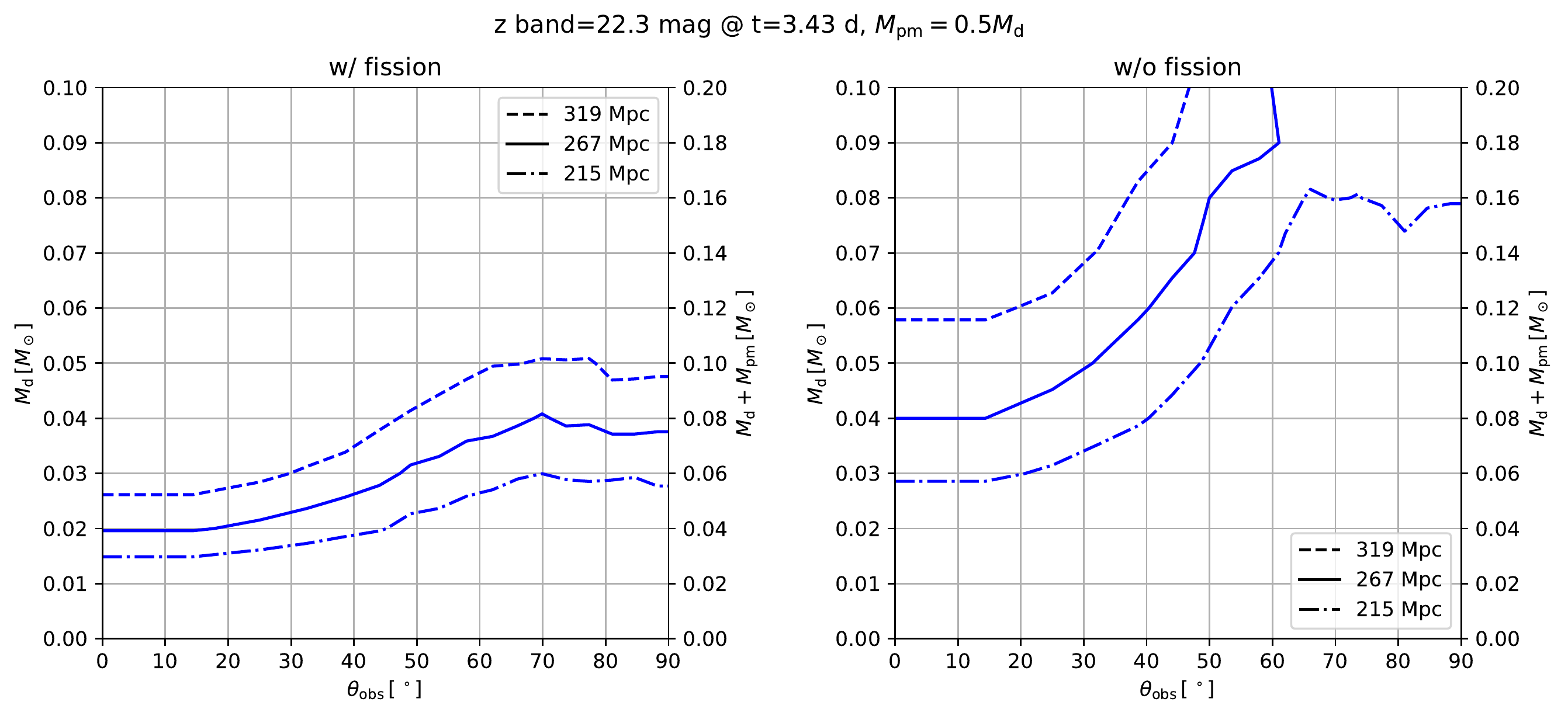}\\
 	 \includegraphics[width=.95\linewidth]{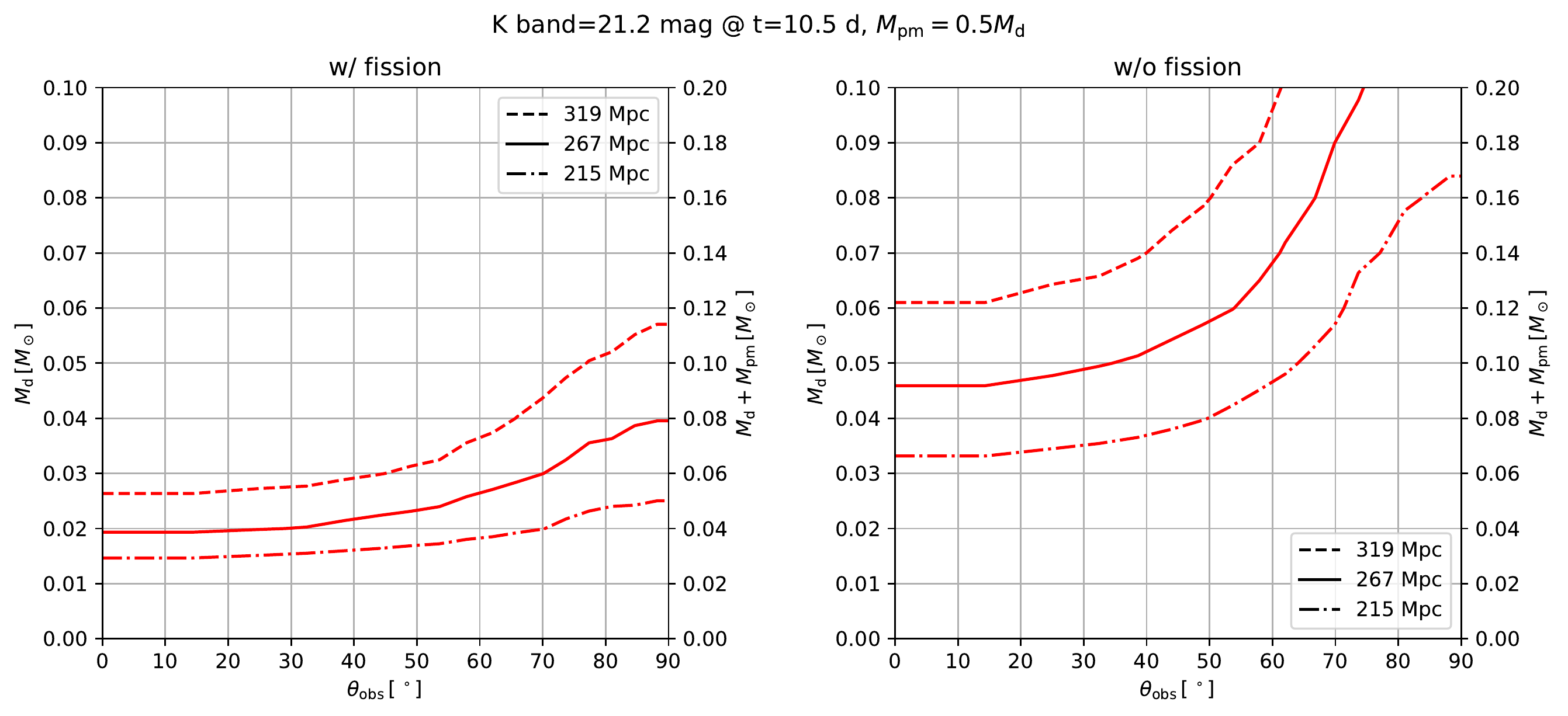}
	 \caption{The same as Figure~\ref{fig:mag_th-meje} but for the models with $M_{\rm pm}=0.5\,M_{\rm d}$.}
	 \label{fig:mag_th-meje-2}
\end{figure*}

Figures~\ref{fig:mag_th-meje-2} shows the upper limit to the dynamical ejecta mass as a function of $\theta_{\rm obs}$ for the models with $M_{\rm pm}=0.5\,M_{\rm d}$. The dynamical ejecta mass larger by $\approx 10$--$50\%$ is allowed for the models with $M_{\rm pm}=0.5\,M_{\rm d}$ than with $M_{\rm pm}=\,M_{\rm d}$ for $\theta_{\rm obs}\leq60^\circ$, while approximately the same upper limits are obtained for $\theta_{\rm obs}\geq70^\circ$. 

\section{High $Y_e$ post-merger ejecta}\label{apx:YHcomp}

\begin{figure*}
 	 \includegraphics[width=.5\linewidth]{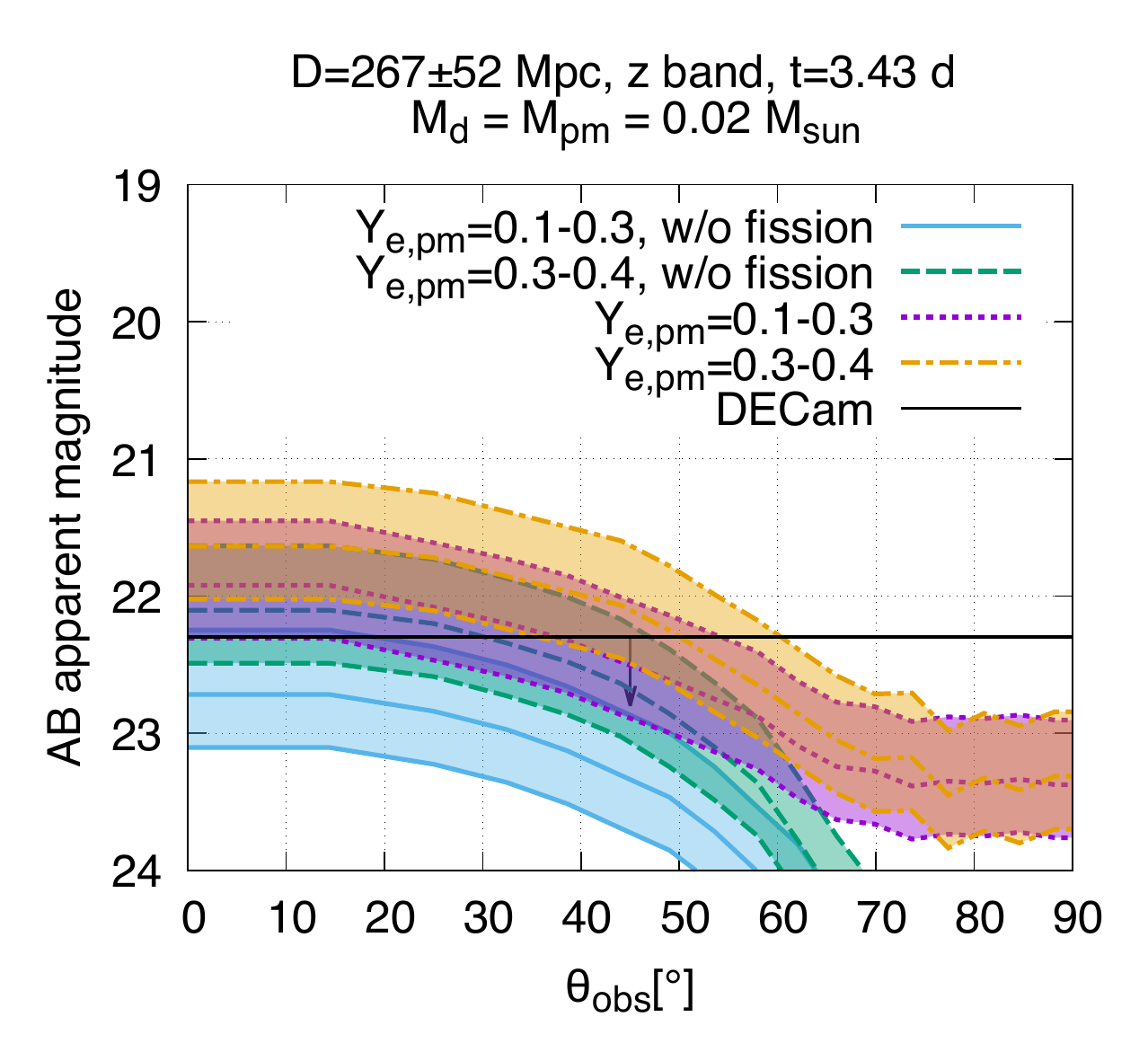}
 	 \includegraphics[width=.5\linewidth]{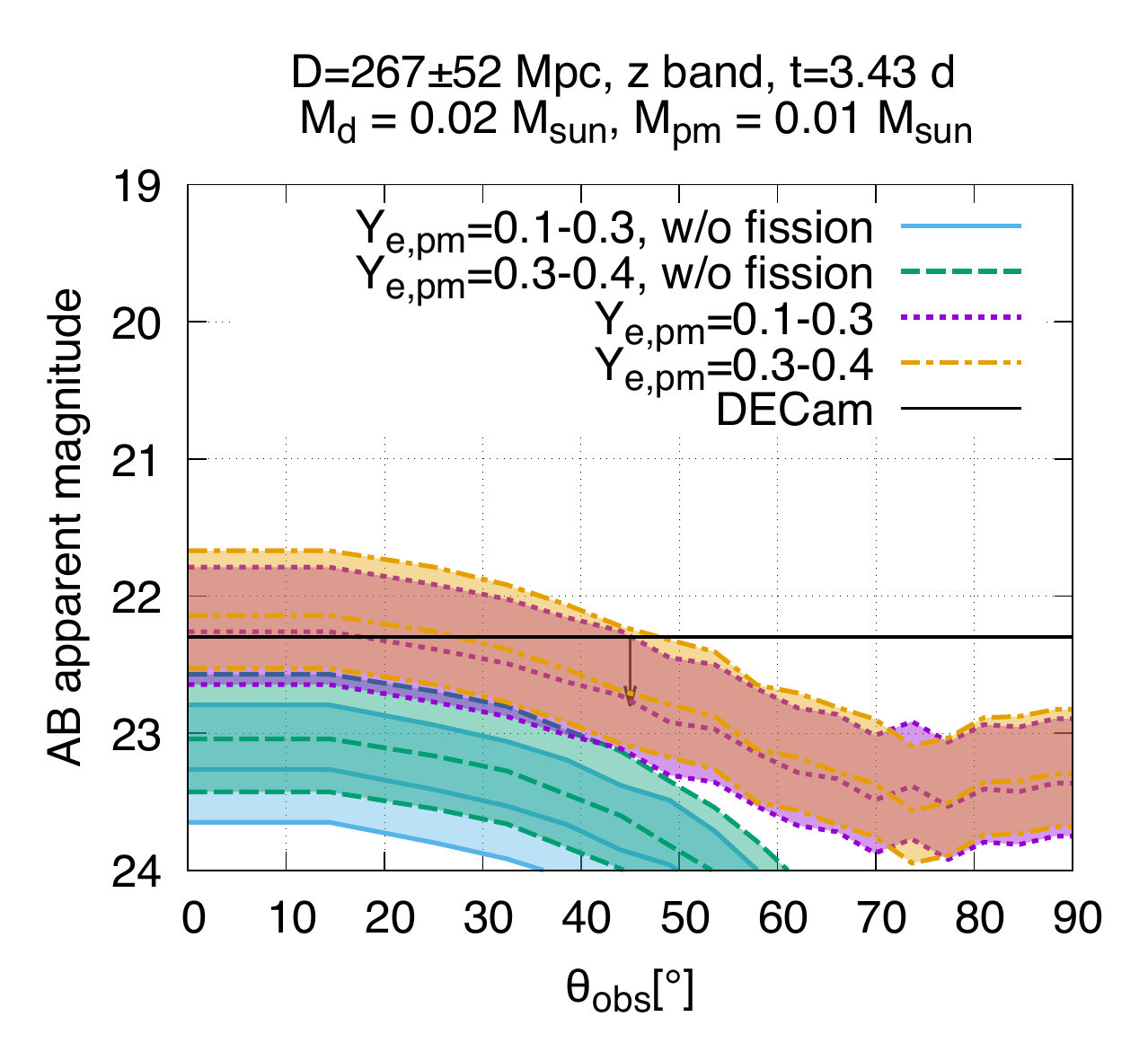}
	 \caption{Comparison of the brightness of the {\it z}-band emission at $t=3.43\,{\rm d}$ between the models with flat $Y_e$ distribution of the post-merger ejecta in $0.1$--$0.3$ and $0.3$--$0.4$. The left and right panels show the models with $(M_{\rm d},M_{\rm pm})=(0.02\,M_\odot, 0.02\,M_\odot)$ and $(0.01\,M_\odot, 0.02\,M_\odot)$, respectively. Curves labeled with "w/o fission" denote the results for the models in which contribution from the fission fragments to the heating rate is omitted. The black horizontal lines in the left and right plots show the upper limits to the {\it z}-band emission at 3.43 d for S190814bv obtained by DECam~\citep{Andreoni:2019qgh}.}
	 \label{fig:mag_YHcomp}
\end{figure*}
Figure~\ref{fig:mag_YHcomp} compares the brightness of the {\it z}-band emission at $t=3.43\,{\rm d}$ between the models with the lanthanide-rich ($Y_e=0.1$--$0.3$) and lanthanide-poor ($Y_e=0.3$--$0.4$) post-merger ejecta. A tighter upper limit to the ejecta mass is obtained for the lanthanide-poor ($Y_e=0.3$--$0.4$) models than the lanthanide-rich ($Y_e=0.1$--$0.3$) models. For the models with $M_{\rm pm}=M_{\rm d}$, the emission for the model with the lanthanide-poor ($Y_e=0.3$--$0.4$) is brighter than $\approx0.5\,{\rm mag}$ than that with the lanthanide-rich ($Y_e=0.1$--$0.3$) post-merger ejecta due to the low value of opacity~\citep{Tanaka:2019iqp} for $\theta_{\rm obs}\lesssim45^\circ$. On the other hand, the enhancement of the emission is less significant for the models with $M_{\rm pm}=0.5\,M_{\rm d}$ due to more significant contribution of the emission from the dynamical ejecta. For both cases with $M_{\rm pm}=M_{\rm d}$ and $M_{\rm pm}=0.5\,M_{\rm d}$, the emission observed from $\theta_{\rm obs}\gtrsim70^\circ$ is approximately identical between the models with lanthanide-rich and lanthanide-poor post-merger ejecta because the emission is dominated by that from the dynamical ejecta.

\end{document}